\documentclass[12pt]{article}

\usepackage{graphicx}
\usepackage{float}
\usepackage{cite}

\def\gtwid{\mathrel{\raise.3ex\hbox{$>$\kern-.75em\lower1ex\hbox{$\sim$}}}}
\def\ltwid{\mathrel{\raise.3ex\hbox{$<$\kern-.75em\lower1ex\hbox{$\sim$}}}}
\def\square{\kern1pt\vbox{\hrule height 1.2pt\hbox{\vrule width 1.2pt\hskip 3pt
   \vbox{\vskip 6pt}\hskip 3pt\vrule width 0.6pt}\hrule height 0.6pt}\kern1pt}

\begin{document}

\begin{titlepage}

\begin{flushright}
UFIFT-QG-20-02 \\
CP3-20-10
\end{flushright}

\vskip 0cm

\begin{center}
{\bf Single Graviton Loop Contribution to the Self-Mass of a Massless, Conformally 
Coupled Scalar on de Sitter Background}
\end{center}

\vskip 0cm

\begin{center}
D. Glavan$^{1*}$, S. P. Miao$^{2\star}$, T. Prokopec$^{3\dagger}$ and
R. P. Woodard$^{4\ddagger}$
\end{center}

\vskip 0cm

\begin{center}
\it{$^{1}$ Centre for Cosmology, Particle Physics and Phenomenology (CP3) \\
Universit\'e catholique de Louvain, Chemin du Cyclotron 2, 1348 Louvain-la-Neuve, BELGIUM}
\end{center}

\begin{center}
\it{$^{2}$ Department of Physics, National Cheng Kung University \\
No. 1 University Road, Tainan City 70101, TAIWAN}
\end{center}

\begin{center}
\it{$^{3}$ Institute for Theoretical Physics, Spinoza Institute \& EMME$\Phi$ \\
Utrecht University, Postbus 80.195, 3508 TD Utrecht, THE NETHERLANDS}
\end{center}

\begin{center}
\it{$^{4}$ Department of Physics, University of Florida,\\
Gainesville, FL 32611, UNITED STATES}
\end{center}

\vspace{0cm}

\begin{center}
ABSTRACT
\end{center}
We use a simplified formalism to re-compute the single graviton loop contribution
to the self-mass of a massless, conformally coupled scalar on de Sitter background 
which was originally made by Boran, Kahya and Park \cite{Boran:2014xpa,Boran:2017fsx,
Boran:2017cfj}. Our result resolves the problem with the flat space correspondence 
limit that was pointed out by Fr\"ob \cite{Frob:2017apy}. We discuss how this 
computation will be used in a long-term project to purge the linearized effective
field equation of gauge dependence.

\begin{flushleft}
\end{flushleft}

\vskip 0cm

\begin{flushleft}
$^{*}$ e-mail: drazen.glavan@uclouvain.be \\
$^{\star}$ e-mail: spmiao5@mail.ncku.edu.tw \\
$^{\dagger}$ e-mail: T.Prokopec@uu.nl \\
$^{\ddagger}$ e-mail: woodard@phys.ufl.edu
\end{flushleft}

\end{titlepage}

\section{Introduction} \label{intro}

Computing fully regulated and renormalized quantum gravitational loop corrections 
is not easy even on flat space background, and is especially challenging on curved 
backgrounds. However, the discovery of a relatively simple gauge \cite{Tsamis:1992xa,
Woodard:2004ut} for de Sitter background has facilitated computations of the 
graviton self-energy \cite{Tsamis:1996qk}, the self-energy of massless \cite{Miao:2005am}
and massive \cite{Miao:2012bj} fermions, the self-mass of a massless, minimally coupled 
\cite{Kahya:2007bc} and conformally coupled \cite{Boran:2014xpa,Boran:2017fsx} scalars
and the vacuum polarization \cite{Leonard:2013xsa}. These 1PI (one-particle irreducible)
2-point functions can be used to quantum-correct the linearized effective field 
equations to infer loop corrections to forces and mode functions using the 
Schwinger-Keldysh formalism \cite{Ford:2004wc}. For example, the vacuum polarization
$i\Bigl[\mbox{}^{\mu} \Pi^{\nu}\Bigr](x;x')$ changes Maxwell's equation in background
metric $g_{\mu\nu}$ to \cite{Prokopec:2002jn,Prokopec:2003bx},
\begin{equation}
\partial_{\nu} \Bigl[ \sqrt{-g} \, g^{\nu\rho} g^{\mu\sigma} F_{\rho\sigma}(x)\Bigr]
+ \int \!\! d^4x' \, \Bigl[\mbox{}^{\mu} \Pi^{\nu}\Bigr](x;x') A_{\nu}(x') = J^{\mu}(x)
\; , \label{QMax}
\end{equation}
where $J^{\mu}(x)$ is the current density, $A_{\mu}$ is the vector potential, and 
$F_{\mu\nu} \equiv \partial_{\mu} A_{\nu} - \partial_{\nu} A_{\mu}$ is the field 
strength. Linearized effective field equations have been solved for one loop corrections
to the graviton mode function (in Hartree approximation) \cite{Mora:2013ypa}, to the
massless fermion mode function \cite{Miao:2006gj}, to the mode functions of minimally 
coupled \cite{Kahya:2007cm} and conformally coupled \cite{Boran:2017cfj} scalars, to the 
photon mode function \cite{Wang:2014tza}, and to electromagnetic forces 
\cite{Glavan:2013jca}.

The results of these studies are fascinating because they often show large logarithmic
corrections to mode functions and exchange potentials. So although the inflationary 
loop counting parameter of $G H^2 \sim 10^{-10}$ is small, the large logarithms can 
make quantum corrections become arbitrarily large at late times and long distances.
However, a note of caution arises from the observation that any graviton loop 
correction is liable to depend upon the gauge fixing procedure. This can be seen 
explicitly in the ability to make flat space corrections vary from plus infinity to
minus infinity \cite{Leonard:2012fs}. Although the flat space gauge dependence must
persist in de Sitter results which survive taking the Hubble parameter to zero, they
might have been absent from the uniquely de Sitter effects responsible for the large 
logarithms \cite{Miao:2012xc}. However, the one time this was checked by making the 
vastly more difficult computation of the vacuum polarization in a different gauge 
\cite{Glavan:2015ura}, it was found that the the coefficient of the large logarithmic 
correction to the photon field strength does change \cite{Glavan:2016bvp}.

A way forward is provided by the recent insight that gauge dependence in the effective
field equations arises from having ignored quantum gravitational corrections from the
source which excites the effective field and from the observer who measures it
\cite{Miao:2017feh}. It is possible to interpret these source and observer effects as
corrections to the 1PI 2-point function using a series of relations derived by Donoghue 
\cite{Donoghue:1994dn,Donoghue:1996mt}. When this is done on flat space all dependence
on the 2-parameter family of Poincar\'e invariant gauges drops out of the linearized 
effective field equation for a minimally coupled scalar \cite{Miao:2017feh}, and the 
equation is presumably completely independent of the gauge fixing procedure. We seek to
include source and observer corrections to 1PI 2-point functions computed on de Sitter 
background. The tensor analysis can be simplified by working with a scalar, however, 
the minimally coupled case is undesirable for two reasons. First, as one can see from
the electromagnetic analog (\ref{QMax}), one loop corrections are sourced by the 
integral of classical solutions against the 1PI 2-point function, and neither the mode 
function \cite{Lifshitz:1945du} nor the exchange potential \cite{Glavan:2019yfc} is 
especially simple for the massless, minimally coupled scalar. Second, it is already 
known that there are no logarithmic enhancements to the massless, minimally coupled 
scalar mode function because gravity only couples to the rapidly redshifting kinetic 
energy \cite{Kahya:2007cm}. 

Getting significant corrections seems to require gravitational couplings such as spin 
that do not redshift \cite{Miao:2008sp}. For scalars that implies either a conformal 
coupling or a mass. The many simplifications associated with the massless, conformally 
coupled scalar support selecting this system for the first analysis of source and 
observer corrections on de Sitter background. The uncorrected 1PI 2-point function has 
been computed by Boran, Kahya and Park \cite{Boran:2014xpa,Boran:2017fsx}, but their 
result was complicated by the decision to express it using de Sitter covariant inverse 
differential operators.\footnote{This decision was made out of deference to concerns 
over the breaking of de Sitter invariance by the graviton propagator which no longer 
seem to be an issue \cite{Higuchi:2001uv,Miao:2009hb,Miao:2011fc,Faizal:2011iv,
Higuchi:2011vw,Miao:2011ng,Morrison:2013rqa,Miao:2013isa,Frob:2014fqa,Woodard:2015kqa}.}
Fr\"ob has also reported a problem with the flat space limit \cite{Frob:2017apy}. For
these reasons we will here carry out an independent computation, taking full advantage
of the scalar's conformal invariance to derive a considerably simpler result with the
correct flat space correspondence limit. Section 2 presents the relevant Feynman rules.
The primitive one graviton contribution is computed in section 3, and its renormalization
is accomplished in section 4. Section 5 discusses the role of this result in our program
of deriving gauge-independent corrections to the scalar mode function and exchange
potential. 

\section{Feynman Rules} \label{Feynman}

The purpose of this section is to present the Feynman rules necessary for
our calculation. We first give the action, both in terms of the original
and conformally rescaled fields. Then we express the self-mass as a a free
expectation value of first and second variations of the action. The various
interactions are somewhat simplified by using the trace-reversed graviton
field. We close by reviewing the scalar and graviton propagators.

The bare Lagrangian for gravity ($D$-dimensional spacelike metric $g_{\mu\nu}$) 
plus a massless, conformally coupled scalar $\phi$ is,
\begin{equation}
\mathcal{L} = \frac{[R \!-\! (D\!-\!2) (D\!-\!1) H^2] \sqrt{-g}}{16 \pi G}
-\frac12 \partial_{\mu} \phi \partial_{\nu} \phi g^{\mu\nu} \sqrt{-g}
- \frac18 \Bigl(\frac{D \!-\! 2}{D \!-\! 1}\Bigr) \phi^2 R \sqrt{-g} \; , 
\label{L}
\end{equation}
where $G$ is Newton's constant and $H$ is the Hubble constant. The background
geometry is,
\begin{equation}
ds^2 = a^2 \Bigl[ -d\eta^2 + d\vec{x} \!\cdot\! d\vec{x} \Bigr] \qquad ,
\qquad a = -\frac1{H\eta} \; , \label{background}
\end{equation}
where the $D-1$ spatial coordinates can be any real number $-\infty < x^i < 
+\infty$, but the conformal time is negative definite $-\infty < \eta < 0$. 
Our work is dramatically simplified by conformally rescaling the metric and the 
scalar,
\begin{equation}
g_{\mu\nu} \equiv a^2 \widetilde{g}_{\mu\nu} \equiv a^2 \Bigl( \eta_{\mu\nu} +
\kappa h_{\mu\nu}\Bigr) \quad , \quad \phi \equiv \frac{\widetilde{\phi}}{
a^{\frac{D}2 -1}} \; , \label{conformal}
\end{equation}
where $\kappa^2 \equiv 16 \pi G$ and graviton indices are raised and lowered 
with the Minkowski metric, $h^{\mu}_{~\nu} \equiv \eta^{\mu\rho} h_{\rho\nu}$.
Up to a surface term the conformal rescaling allows us to express the Lagrangian 
as,
\begin{eqnarray}
\lefteqn{\mathcal{L} \longrightarrow \frac{(D\!-\!2)}{2} H a^{D-1} 
\sqrt{-\widetilde{g}} \, \widetilde{g}^{\rho\sigma} \widetilde{g}^{\mu\nu} 
h_{\rho\sigma ,\mu} h_{\nu 0} + a^{D-2} \sqrt{-\widetilde{g}} \, 
\widetilde{g}^{\alpha\beta} \widetilde{g}^{\rho\sigma} \widetilde{g}^{\mu\nu} }
\nonumber \\
& & \hspace{1cm} \times \Biggl\{ \frac12 h_{\alpha\rho ,\mu} h_{\nu\sigma ,\beta} 
\!-\! \frac12 h_{\alpha\beta ,\rho} h_{\sigma\mu ,\nu} \!+\! \frac14 
h_{\alpha\beta ,\rho} h_{\mu\nu ,\sigma} \!-\! \frac14 h_{\alpha\rho ,\mu} 
h_{\beta\sigma ,\nu} \Biggr\} \nonumber \qquad \\
& & \hspace{4.5cm} -\frac12 \partial_{\mu} \widetilde{\phi} \partial_{\nu} 
\widetilde{\phi} \widetilde{g}^{\mu\nu} \sqrt{-\widetilde{g}} - \frac18 \Bigl(
\frac{D \!-\! 2}{D \!-\! 1}\Bigr) \widetilde{\phi}^2 \widetilde{R} 
\sqrt{-\widetilde{g}} \; . \qquad \label{Lconf}
\end{eqnarray}
Note that $\widetilde{g}^{\alpha\beta}$ and $\widetilde{R}$ are infinite order in
the graviton field.

\begin{figure}[H]
\includegraphics[width=13cm,height=3cm]{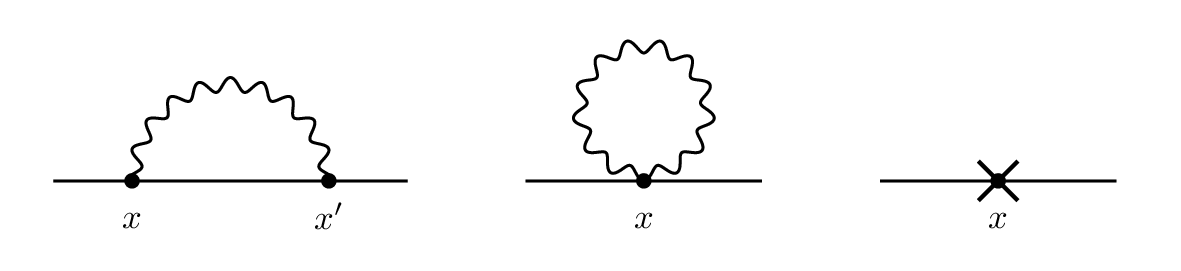}
\caption{\footnotesize{One graviton contributions to $-i M^2(x;x')$. Graviton lines 
are wavy and scalar lines are straight. Counterterms are denoted by a cross.}} 
\label{diagrams}
\end{figure}

The single graviton loop contribution to the self-mass of $\widetilde{\phi}$ 
can be expressed in terms of free expectation values of variations of the
action,
\begin{equation}
-i \widetilde{M}^2(x;x') = \Biggl\langle \Omega \Biggl\vert T^*\Biggl\{ 
\Biggl[ \frac{i \delta S}{\delta \widetilde{\phi}(x)} \Biggr]_{h \widetilde{\phi}} 
\times \Biggl[ \frac{i \delta S}{\delta \widetilde{\phi}(x')}\Biggr]_{h 
\widetilde{\phi}} + \Biggl[ \frac{i \delta^2 S}{\delta \widetilde{\phi}(x) 
\delta \widetilde{\phi}(x')} \Biggr]_{h h} \Biggr\} \Biggr\vert \Omega 
\Biggr\rangle \; , \label{M2start}
\end{equation}
where the $T^*$-ordering symbol indicates that any derivative is acted outside
the time-ordered product of the fields. The self-mass of the original field 
$\phi$ is easy to recover,
\begin{equation}
-i M^2(x;x') \equiv (a a')^{\frac{D}2 -1} \times -i \widetilde{M}^2(x;x') \; .
\end{equation}
The first term in (\ref{M2start}) corresponds to the left hand diagram of 
Figure~\ref{diagrams} while the second term is represented by the middle diagram.

The variations in expression (\ref{M2start}) are,
\begin{eqnarray}
\frac{\delta S}{\delta \widetilde{\phi}(x)} & \!\!\!=\!\!\! & \partial_{\mu} \Bigl( 
\sqrt{-\widetilde{g}} \, \widetilde{g}^{\mu\nu} \partial_{\nu} \widetilde{\phi}\Bigr) 
-\frac14 \Bigl( \frac{D \!-\! 2}{D \!-\! 1}\Bigr) \widetilde{\phi} \widetilde{R}
\sqrt{-\widetilde{g}}  \; , \label{dS/dphi} \\
\frac{\delta^2 S}{\delta \widetilde{\phi}(x) \delta \widetilde{\phi}(x')} 
& \!\!\!=\!\!\! & \partial_{\mu} \Bigl( \sqrt{-\widetilde{g}} \, \widetilde{g}^{\mu\nu} 
\partial_{\nu} \delta^D(x \!-\! x') \Bigr) - \frac14 \Bigl( \frac{D \!-\! 2}{D \!-\! 1}\Bigr) 
\widetilde{R} \sqrt{-\widetilde{g}} \delta^D(x \!-\! x') \; . \quad \label{d2S/dphi2} 
\end{eqnarray}
It is best to break (\ref{dS/dphi}) up into two parts,
\begin{equation}
\frac{\delta S_a}{\delta \widetilde{\phi}(x)} = \partial_{\mu} \Bigl( 
\sqrt{-\widetilde{g}} \, \widetilde{g}^{\mu\nu} \partial_{\nu} \widetilde{\phi}\Bigr) 
\qquad , \qquad \frac{\delta S_b}{\delta \widetilde{\phi}(x)} = -\frac14 \Bigl( 
\frac{D \!-\! 2}{D \!-\! 1}\Bigr) \widetilde{\phi} \widetilde{R} \sqrt{-\widetilde{g}} 
\; , \label{Sab}
\end{equation}
and similarly for the 2nd variational derivatives. The variation of $S_b$ can be 
usefully further decomposed with the identity,
\begin{eqnarray}
\lefteqn{\sqrt{-\widetilde{g}} \widetilde{R} = \partial_{\rho} \Bigl[ 
\sqrt{-\widetilde{g}} \Bigl( \widetilde{g}^{\mu\nu} \widetilde{\Gamma}^{\rho}_{~\mu\nu}
- \widetilde{g}^{\rho\mu} \widetilde{\Gamma}^{\nu}_{~\nu \mu}\Bigr) \Bigr] } \nonumber \\
& & \hspace{6cm} + \sqrt{-\widetilde{g}} \, \widetilde{g}^{\mu\nu} \Bigl(
\widetilde{\Gamma}^{\rho}_{~\sigma\mu} \widetilde{\Gamma}^{\sigma}_{~\rho\nu} -
\widetilde{\Gamma}^{\rho}_{~\rho \sigma} \widetilde{\Gamma}^{\sigma}_{~\mu\nu} 
\Bigr) \; . \qquad
\end{eqnarray}
Hence we have,
\begin{eqnarray}
\frac{\delta S_{b1}}{\delta \widetilde{\phi}(x)} & = & -\frac14 \Bigl( 
\frac{D \!-\! 2}{D \!-\! 1}\Bigr) \widetilde{\phi} \partial_{\rho} \Bigl[ 
\sqrt{-\widetilde{g}} \Bigl( \widetilde{g}^{\mu\nu} \widetilde{\Gamma}^{\rho}_{~\mu\nu}
- \widetilde{g}^{\rho\mu} \widetilde{\Gamma}^{\nu}_{~\nu \mu}\Bigr) \Bigr] \; , \qquad \\
\frac{\delta S_{b2}}{\delta \widetilde{\phi}(x)} & = & -\frac14 \Bigl( 
\frac{D \!-\! 2}{D \!-\! 1}\Bigr) \widetilde{\phi} \sqrt{-\widetilde{g}} \, 
\widetilde{g}^{\mu\nu} \Bigl(\widetilde{\Gamma}^{\rho}_{~\sigma\mu} 
\widetilde{\Gamma}^{\sigma}_{~\rho\nu} - \widetilde{\Gamma}^{\rho}_{~\rho \sigma} 
\widetilde{\Gamma}^{\sigma}_{~\mu\nu} \Bigr) \; . \qquad
\end{eqnarray}
The expansions of all variations in powers of the gravitons are simplified using the 
trace-reversed graviton field,
\begin{equation}
\widehat{h}_{\mu\nu} \equiv h_{\mu\nu} - \frac12 \eta_{\mu\nu} h \; .
\end{equation}
The expansion of the two first variations are,
\begin{eqnarray}
\Biggl[ \frac{i \delta S_a}{\delta \widetilde{\phi}(x)}\Biggr]_{h \widetilde{\phi}} 
& = & -i\kappa \partial_{\mu} \Bigl[\widehat{h}^{\mu\nu} \partial_{\nu} 
\widetilde{\phi} \Bigr] \; , \label{a} \\
\Biggl[ \frac{i \delta S_b}{\delta \widetilde{\phi}(x)}\Biggr]_{h \widetilde{\phi}} 
& = & -\frac{i \kappa}{4} \Bigl( \frac{D \!-\! 2}{D \!-\! 1}\Bigr) \widetilde{\phi} 
\Bigl[ \partial_{\mu} \partial_{\nu} + \frac{\eta_{\mu\nu} \partial^2}{D \!-\! 2} 
\Bigr] \widehat{h}^{\mu\nu} \; . \label{b}
\end{eqnarray}
The second variations are simplest when expressed using both $h_{\mu\nu}$ and
$\widehat{h}_{\mu\nu}$,
\begin{eqnarray}
\Biggl[ \frac{i \delta^2 S_a}{\delta \widetilde{\phi}(x) \delta \widetilde{\phi}(x')}
\Biggr]_{h h} \!\!\!\!\!\!\!\!\!\!&=&\!\!\!\! i\kappa^2 \partial_{\mu} 
\Biggl[\Bigl[\widehat{h}^{\mu\rho} h^{\nu}_{~\rho} \!-\! \frac14 \eta^{\mu\nu} 
\widehat{h}^{\rho\sigma} h_{\rho\sigma} \Bigr] \partial_{\nu} \delta^D(x \!-\! x') 
\Biggr] , \quad \label{A} \\
\Biggl[\frac{i \delta^2 S_{b1}}{\delta \widetilde{\phi}(x) \delta \widetilde{\phi}(x')}
\Biggr]_{h h} \!\!\!\!\!\!\!\!\!\!&=&\!\!\!\! \frac{i\kappa^2}4 \Bigl( 
\frac{D \!-\!2}{D\!-\!1}\Bigr) \nonumber \\
& & \hspace{-0.5cm} \times \partial_{\rho} \Biggl[ \partial^{\alpha} \Bigl(
h^{\rho \beta} \widehat{h}_{\alpha\beta} \Bigr) - \frac12 \partial^{\rho} 
\Bigl( h^{\alpha\beta} \widehat{h}_{\alpha\beta} \Bigr) - \frac12 
h^{\rho\sigma} h_{,\sigma}\Biggr] \delta^D(x \!-\! x') \; , \qquad 
\label{B1} \\
\Biggl[\frac{i \delta^2 S_{b2}}{\delta \widetilde{\phi}(x) \delta \widetilde{\phi}(x')}
\Biggr]_{h h} \!\!\!\!\!\!\!\!\!\!&=&\!\!\!\! \frac{i\kappa^2}4 \Bigl( 
\frac{D \!-\!2}{D\!-\!1}\Bigr) \Biggl[ -\frac12 \widehat{h}^{\rho\sigma ,\mu} 
\widehat{h}_{\mu\rho ,\sigma} \!+\! \frac14 \widehat{h}^{\rho\sigma ,\mu} 
h_{\rho\sigma , \mu} \Biggr] \delta^D(x\!-\!x') \; . \quad \label{B2} 
\end{eqnarray}
Note that the two graviton fields in each of these expressions are both evaluated
at the point $x^{\mu}$.

The propagator of $\widetilde{\phi}$ is the same as for a massless scalar in flat space,
\begin{equation}
i\Delta(x;x') = \frac{\Gamma(\frac{D}2 \!-\! 1)}{4 \pi^{\frac{D}2} \Delta x^{D-2}}
\qquad , \qquad \Delta x^2(x;x') \equiv \Bigl\Vert \vec{x} \!-\! \vec{x}' \Bigr\Vert^2
- \Bigl( \vert \eta \!-\! \eta'\vert - i \epsilon\Bigr)^2 \; . \label{propphi}
\end{equation}
The graviton propagator is defined by adding the gauge fixing term \cite{Tsamis:1992xa,
Woodard:2004ut},
\begin{equation}
\mathcal{L}_{\rm GF} = -\frac12 a^{D-2} \eta^{\mu\nu} F_{\mu} F_{\nu} \quad , 
\quad F_{\mu} = \eta^{\rho\sigma} \Bigl( h_{\mu\rho , \sigma} - \frac12 
h_{\rho\sigma ,\mu} + (D \!-\! 2) H a h_{\mu\rho} \delta^0_{~\sigma}\Bigr) .
\label{gauge}
\end{equation}
The graviton propagator in this gauge takes the form,
\begin{equation}
i\Bigl[\mbox{}_{\mu\nu} \Delta_{\rho\sigma}\Bigr](x;x') = \sum_{I=A,B,C} 
\Bigl[ \mbox{}_{\mu\nu} T^I_{\rho\sigma}\Bigr] \times i \Delta_I(x;x') \; .
\label{proph}
\end{equation}
The three propagators $i\Delta_I(x;x')$ are for minimally coupled scalars with
masses $M^2_A = 0$, $M^2_B = (D-2) H^2$ and $M^2_C = 2(D-3) H^2$, so our four
scalar propagators obey the equations,
\begin{equation}
\partial^2 i\Delta(x;x') = i\delta^D(x \!-\! x') \; , \; \Bigl[ \partial^2 
- (D\!-\!2) H a \partial_0 - M^2_I a^2\Bigr] i\Delta_I(x;x') = \frac{i 
\delta^D(x \!-\! x')}{a^{D-2}} \; . \label{propeqns}
\end{equation}
Because $-i \widetilde{M}^2(x;x')$ is quarticly divergent at one loop we only
need the first few terms in the expansions of the three $i\Delta_I(x;x')$
\cite{Tsamis:1992xa,Woodard:2004ut,Onemli:2004mb},
\begin{eqnarray}
\lefteqn{ i\Delta_A = \frac{i\Delta(x;x')}{(aa')^{\frac{D}2 -1}} + 
\frac{\Gamma(\frac{D}2 \!+\! 1)}{8 \pi^{\frac{D}2} (D\!-\!4)}
\frac{H^2}{(a a' \Delta x^2)^{\frac{D}2 -2}} + 
\frac{\Gamma(\frac{D}2 \!+\! 2)}{64 \pi^{\frac{D}2} (D\!-\! 6)} \frac{H^4}{(a a' 
\Delta x^2)^{\frac{D}2 -3}} } \nonumber \\
& & \hspace{-0.7cm} - \frac{H^{D-2}}{(4 \pi)^{\frac{D}2}} \frac{\Gamma(D \!-\! 1)}{
\Gamma(\frac{D}2)} \Bigl[ \pi {\rm cot}\Bigl( \frac{\pi D}{2}\Bigr) \!-\! \ln(a a')
\Bigr] + \frac{H^{D}}{(4 \pi)^{\frac{D}2}} \frac{\Gamma(D)}{\Gamma(\frac{D}2 \!+\! 1)}
\frac{a a' \Delta x^2}{4} + \cdots , \qquad \label{propA} \\
\lefteqn{ i\Delta_B = \frac{i\Delta(x;x')}{(aa')^{\frac{D}2 -1}} + 
\frac{\Gamma(\frac{D}2)}{16 \pi^{\frac{D}2}} \frac{H^2}{(a a' \Delta x^2)^{\frac{D}2 -2}}
+ \frac{\Gamma(\frac{D}2 \!+\! 1)}{128 \pi^{\frac{D}2}} \frac{H^4}{(a a' \Delta x^2)^{
\frac{D}2 -3}} } \nonumber \\
& & \hspace{2.5cm} - \frac{H^{D-2}}{(4 \pi)^{\frac{D}2}} \frac{\Gamma(D \!-\! 2)}{
\Gamma(\frac{D}2)} - \frac{H^{D}}{(4 \pi)^{\frac{D}2}} \frac{\Gamma(D\!-\! 1)}{
\Gamma(\frac{D}2 \!+\! 1)} \frac{a a' \Delta x^2}{4} + \cdots \; , \qquad 
\label{propB} \\
\lefteqn{ i\Delta_C \!=\! \frac{i\Delta(x;x')}{(aa')^{\frac{D}2 -1}} \!+\! 
\frac{(D \!-\! 6) \Gamma(\frac{D}2 \!-\! 1)}{32 \pi^{\frac{D}2}}
\frac{H^2}{(a a' \Delta x^2)^{\frac{D}2 -2}} \!+\! 
\frac{(D \!-\! 8) \Gamma(\frac{D}2)}{256 \pi^{\frac{D}2}} \frac{H^4}{(a a' \Delta x^2)^{
\frac{D}2 -3}} } \nonumber \\
& & \hspace{2.5cm} + \frac{H^{D-2}}{(4 \pi)^{\frac{D}2}} \frac{\Gamma(D \!-\! 3)}{
\Gamma(\frac{D}2)} + \frac{H^{D}}{(4 \pi)^{\frac{D}2}} \frac{\Gamma(D \!-\! 2)}{
\Gamma(\frac{D}2 \!+\! 1)} \frac{a a' \Delta x^2}{2} + \cdots \qquad \label{propC}
\end{eqnarray}

The three tensor factors of (\ref{proph}) are constructed from $\eta_{\mu\nu}$, 
$\overline{\eta}_{\mu\nu} \equiv \eta_{\mu\nu} + \delta^0_{~\mu} \delta^0_{~\nu}$ 
and $\delta^0_{~\mu}$,
\begin{eqnarray}
\Bigl[\mbox{}_{\mu\nu} T^A_{\rho\sigma}\Bigr] & = & 2 \, \overline{\eta}_{\mu (\rho} 
\overline{\eta}_{\sigma) \nu} - \frac{2}{D \!-\! 3} \, \overline{\eta}_{\mu\nu}
\overline{\eta}_{\rho\sigma} \quad , \quad \Bigl[\mbox{}_{\mu\nu} T^B_{\rho\sigma}
\Bigr] = -4 \delta^0_{~(\mu} \overline{\eta}_{\nu ) (\rho} \delta^0_{~\sigma)} 
\; , \qquad \label{TAB} \\
\Bigl[\mbox{}_{\mu\nu} T^C_{\rho\sigma}\Bigr] & = & \frac{2 E_{\mu\nu}
E_{\rho\sigma}}{(D \!-\! 2) (D \!-\! 3)} \quad , \quad E_{\mu\nu} \equiv 
(D \!-\!3) \delta^0_{~\mu} \delta^0_{~\nu} \!+\! \overline{\eta}_{\mu\nu} 
\; , \label{TC}  
\end{eqnarray}
where parenthesized indices are symmetrized. In addition to reducing the number
of terms in each interaction, the trace-reversed graviton $\widehat{h}_{\mu\nu}$
has a simpler propagator than $h_{\mu\nu}$. The mixed propagator is,
\begin{eqnarray}
\lefteqn{ i\Bigl[\mbox{}_{\widehat{\mu\nu}} \Delta_{\rho\sigma}\Bigr] \equiv 
\Bigl[\delta^{\alpha}_{\mu} \delta^{\beta}_{\nu} - \frac12 \eta_{\mu\nu} 
\eta^{\alpha\beta} \Bigr] i\Bigl[\mbox{}_{\alpha\beta} \Delta_{\rho\sigma}\Bigr] =
2 \eta_{\mu (\rho} \eta_{\sigma) \nu} i \Delta_A } \nonumber \\
& & \hspace{3cm} + 4 \delta^{0}_{(\mu} \overline{\eta}_{\nu) (\rho} 
\delta^{0}_{\sigma)} \Bigl[ i\Delta_A \!-\! i\Delta_B\Bigr] - 
\frac{2 \delta^{0}_{\mu} \delta^{0}_{\nu} E_{\rho\sigma}}{D \!-\! 3} \Bigl[
i\Delta_A \!-\! i\Delta_C\Bigr] \; . \qquad \label{prophatunhat}
\end{eqnarray}
Note from the propagator expansions (\ref{propA}-\ref{propC}) that the differences
$[i\Delta_A - i\Delta_B]$ and $[i\Delta_A - i\Delta_C]$ are less singular near
coincidence than any of the $i\Delta_I$. The fully trace-reversed propagator is, 
\begin{eqnarray}
\lefteqn{ i\Bigl[\mbox{}_{\widehat{\mu\nu}} \Delta_{\widehat{\rho\sigma}}\Bigr] 
\equiv \Bigl[\delta^{\alpha}_{\rho} \delta^{\beta}_{\sigma} - \frac12 
\eta_{\rho\sigma} \eta^{\alpha\beta} \Bigr] i\Bigl[\mbox{}_{\widehat{\mu\nu}} 
\Delta_{\alpha\beta}\Bigr] = \Bigl[2 \eta_{\mu (\rho} \eta_{\sigma) \nu} - 
\eta_{\mu\nu} \eta_{\rho\sigma}\Bigr] i \Delta_A } \nonumber \\
& & \hspace{1.5cm} + 4 \delta^{0}_{(\mu} \overline{\eta}_{\nu) (\rho} 
\delta^{0}_{\sigma)} \Bigl[ i\Delta_A \!-\! i\Delta_B\Bigr] - 2 \Bigl(
\frac{D \!-\! 2}{D \!-\! 3}\Bigr) \delta^{0}_{\mu} \delta^{0}_{\nu} \delta^0_{\rho}
\delta^0_{\sigma} \Bigl[i\Delta_A \!-\! i\Delta_C\Bigr] \; . \qquad 
\label{prophathat}
\end{eqnarray}

\section{Primitive Contributions} \label{primitive}

The purpose of this section is to evaluate the left hand and middle diagrams
of Figure~\ref{diagrams}. We first compute the simpler, middle diagram, which 
is based on the interactions (\ref{A}-\ref{B2}). Then we tackle the more difficult, 
middle diagram which is based on products of the interactions (\ref{a}-\ref{b}).
The section closes with a total for all primitive diagrams. 

\subsection{4-Point Contributions} \label{4-point}

The middle diagram of Figure~\ref{diagrams} consists of a single, differentiated 
and coincident graviton propagator evaluated between the interactions (\ref{A}-\ref{B2}).
When one takes the coincidence limit {\it before} differentiating, the only spacetime
dependence is the $\ln(a)$ part of the $i\Delta_A$ propagator,
\begin{eqnarray}
i\Delta_A(x;x') \Bigl\vert_{x'=x} \!\!\!\!\!\!\!\!\!\!\!\!\!& = & \!\!\!- 
\frac{H^{D-2}}{(4 \pi)^{\frac{D}2}} \frac{\Gamma(D \!-\! 1)}{\Gamma(\frac{D}2)} 
\Bigl[ \pi {\rm cot}\Bigl( \frac{\pi D}{2}\Bigr) \!-\! 2 \ln(a) \Bigr] \!\rightarrow 
-A_0 \!+\! \frac{H^2 \ln(a)}{4 \pi^2} , \quad \label{DAcoinc} \\
i\Delta_B(x;x') \Bigl\vert_{x'=x} \!\!\!\!\!\!\!\!\!\!\!\!\!& = &\!\!\! -
\frac{H^{D-2}}{(4\pi)^{\frac{D}2}} \frac{\Gamma(D\!-\!2)}{\Gamma(\frac{D}2)} 
\longrightarrow -\frac{H^2}{16 \pi^2} \equiv B_0 \; , \label{DBcoinc} \\
i\Delta_C(x;x') \Bigl\vert_{x'=x} \!\!\!\!\!\!\!\!\!\!\!\!\!& = &\!\!\! +
\frac{H^{D-2}}{(4\pi)^{\frac{D}2}} \frac{\Gamma(D\!-\!3)}{\Gamma(\frac{D}2)} 
\longrightarrow +\frac{H^2}{16 \pi^2} \equiv C_0 \; . \label{DCcoinc}
\end{eqnarray}
Note that only $A_0$ is divergent. Taking the coincidence limit {\it after} 
differentiating produces only finite results,
\begin{eqnarray}
\partial_{\mu} \partial'_{\nu} i\Delta_A(x;x') \Bigl\vert_{x'=x} & = & -\frac12 g_{\mu\nu}
\!\times\! \frac{H^{D}}{(4 \pi)^{\frac{D}2}} \frac{\Gamma(D)}{\Gamma(\frac{D}2 \!+\! 1)} 
\longrightarrow -\frac{3 H^4}{32 \pi^2} \, g_{\mu\nu} \; , \label{ddDAcoinc} \\
\partial_{\mu} \partial'_{\nu} i\Delta_B(x;x') \Bigl\vert_{x'=x} & = & +\frac12 g_{\mu\nu}
\!\times\! \frac{H^{D}}{(4\pi)^{\frac{D}2}} \frac{\Gamma(D\!-\!1)}{\Gamma(\frac{D}2 \!+\!1)} 
\longrightarrow +\frac{H^4}{32 \pi^2} \, g_{\mu\nu} \; , \label{ddDBcoinc} \\
\partial_{\mu} \partial'_{\nu} i\Delta_C(x;x') \Bigl\vert_{x'=x} & = & -g_{\mu\nu} 
\!\times\! \frac{H^{D}}{(4\pi)^{\frac{D}2}} \frac{\Gamma(D\!-\!2)}{\Gamma(\frac{D}2 \!+\! 1)} 
\longrightarrow -\frac{H^4}{32 \pi^2} \, g_{\mu\nu} \; . \label{ddDCcoinc}
\end{eqnarray}

\subsubsection{$-i \widetilde{M}^2_{A}(x;x')$} \label{MsubA}

What we might call $-i \widetilde{M}^2_A(x;x')$ is the free expectation value of (\ref{A}),
\begin{eqnarray}
\lefteqn{-i \widetilde{M}^2_A = i\kappa^2 \partial_{\mu} \Biggl\langle \!\!\Omega 
\Biggl\vert T^*\Biggl[\widehat{h}^{\mu\rho}(x) h^{\nu}_{~\rho}(x) \!-\! \frac{\eta^{\mu\nu}}{4} 
\widehat{h}^{\rho\sigma}(x) h_{\rho\sigma}(x) \Biggr] \partial_{\nu} \delta^D(x \!-\! x') 
\Biggr\vert \Omega \!\!\Biggr\rangle ,\quad} \label{MA1} \\
& & \hspace{1.4cm} = i\kappa^2 \partial_{\mu} \Biggl\{ \Biggl[ i 
\Bigl[\mbox{}^{\widehat{\mu\rho}} \Delta^{\nu}_{~\rho}\Bigr](x;x) - \frac{\eta^{\mu\nu}}{4} 
i \Bigl[\mbox{}^{\widehat{\rho\sigma}} \Delta_{\rho\sigma}\Bigr](x;x) \Biggr] \partial_{\nu} 
\delta^D(x \!-\! x') \Biggr\} . \qquad \label{MA2}
\end{eqnarray}
The two contracted propagators we require are,
\begin{eqnarray}
i \Bigl[\mbox{}^{\widehat{\mu\rho}} \Delta^{\nu}_{~\rho}\Bigr] \!\!\!\!& = &\!\!\!\!
\eta^{\mu\nu} \Bigl[(D\!-\!1) i\Delta_B \!+\! 2 i\Delta_C\Bigr] + \overline{\eta}^{\mu\nu} 
\Bigl[D i\Delta_A \!-\! (D \!-\!2) i\Delta_B \!-\! 2 i\Delta_C\Bigr] , \qquad 
\label{contract1} \\ 
i \Bigl[\mbox{}^{\widehat{\rho\sigma}} \Delta_{\rho\sigma}\Bigr] \!\!\!\!& = &\!\!\!\! 
D (D \!-\! 1) i\Delta_A + 2 (D\!-\!1) i\Delta_{B} + 2 i\Delta_{C} \; . \qquad
\label{contract2}
\end{eqnarray}
Substituting (\ref{contract1}-\ref{contract2}) in (\ref{MA2}) and making use of 
(\ref{DAcoinc}-\ref{DCcoinc}) gives,
\begin{equation}
-i \widetilde{M}^2_A(x;x') = -\frac{\kappa^2}{4} D (D\!-\!1) A_0 \partial 
\!\cdot\! \partial' \Biggl[ \frac{i \delta^D(x \!-\! x')}{(a a')^{\frac{D}2 -2}} \Biggr]
+ \kappa^2 D A_0 \vec{\nabla} \!\cdot\! \vec{\nabla}' \Biggl[ 
\frac{ i \delta^D(x \!-\!x')}{(a a')^{\frac{D}2 -2}} \Biggr] . \label{finalMA}
\end{equation}

\subsubsection{$-i \widetilde{M}^2_{B}(x;x')$} \label{MsubB}

What should be called $-i \widetilde{M}^2_{B1}(x;x')$ is the free expectation value
of (\ref{B1}),
\begin{eqnarray}
\lefteqn{-i \widetilde{M}^2_{B1} \equiv \Biggl\langle \Omega \Biggl\vert T^*\Biggl[
\frac{i \delta^2 S_{b1}}{\delta \widetilde{\phi}(x) \delta \widetilde{\phi}(x')}
\Biggr]_{h h} \Biggr\vert \Omega \Biggr\rangle = \frac{\kappa^2}{4} \Bigl( 
\frac{D\!-\!2}{D\!-\!1} \Bigr) \partial_{\rho} \Biggl[ \partial^{\sigma} 
i\Bigl[\mbox{}^{\rho\beta} \Delta_{\widehat{\sigma\beta}}\Bigr](x;x) } \nonumber \\
& & \hspace{1.6cm} - \frac12 \partial^{\rho} i\Bigl[ \mbox{}^{\alpha\beta} 
\Delta_{\widehat{\alpha\beta}}\Bigr](x;x) - \frac12 \partial_{\sigma}' 
i\Bigl[\mbox{}^{\rho\sigma} \Delta^{\alpha}_{~\alpha}\Bigr](x;x')
\Bigr\vert_{x' = x} \Biggr] i \delta^D(x \!-\! x') \; . \qquad \label{MB11} 
\end{eqnarray}
Because the three propagators inside the square bracket of (\ref{MB11}) are singly
differentiated, only the $\ln(a)$ part of the $i\Delta_A$ can contribute, and only
when differentiated with respect to time. Hence the first and the third propagators
give nothing, and the contribution from the second propagator is finite,
\begin{equation}
-i \widetilde{M}^2_{B1}(x;x') = \frac{\kappa^2 H^4 a^2}{4 \pi^2} \, 
i\delta^4(x \!-\! x') \; . \label{finalMB1}
\end{equation}

The second 4-point contribution is,
\begin{eqnarray}
\lefteqn{-i \widetilde{M}^2_{B2}(x;x') \equiv \Biggl\langle \Omega \Biggl\vert
T^*\Biggl[\frac{i \delta^2 S_{b1}}{\delta \widetilde{\phi}(x) \delta \widetilde{\phi}(x')}
\Biggr]_{h h} \Biggr\vert \Omega \Biggr\rangle } \nonumber \\
& & \hspace{-0.7cm} = \frac{\kappa^2}{8} \Bigl( \frac{D\!-\!2}{D\!-\!1}\Bigr) \Biggl[ 
-\partial^{\mu} \partial_{\sigma}' i\Bigl[\mbox{}^{\widehat{\rho\sigma}} 
\Delta_{\widehat{\mu\rho}}\Bigr](x;x') + \frac12 \partial \!\cdot\! \partial' 
i\Bigl[\mbox{}^{\widehat{\rho\sigma}} \Delta_{\rho\sigma}\Bigr](x;x') \Biggr]
i \delta^D(x \!-\! x') \; . \qquad \label{MB21}
\end{eqnarray}
By virtue of the delta function, each of the two propagators in (\ref{MB21}) 
is finite and consists of a sum of the three terms (\ref{ddDAcoinc}-\ref{ddDCcoinc}), 
so it remains only to determine the coefficients. Those of the second propagator come
from expression (\ref{contract2}), and the first propagator derives from a single
contraction of (\ref{prophathat}),
\begin{eqnarray}
\lefteqn{i\Bigl[\mbox{}^{\widehat{\rho\sigma}} \Delta_{\widehat{\mu\rho}}\Bigr] =
\delta^{\sigma}_{\mu} \Biggl[ -\Bigl(\frac{D\!-\!1}{D\!-\!3}\Bigr) i\Delta_A + 
(D\!-\!1) i\Delta_B + 2 \Bigl( \frac{D\!-\!2}{D\!-\!3}\Bigr) i\Delta_C\Biggr] } 
\nonumber \\
& & \hspace{3.5cm} + \overline{\delta}^{\sigma}_{\mu} (D \!-\! 2) \Biggl[ \Bigl(
\frac{D\!-\!1}{D\!-\!3}\Bigr) i \Delta_A - i\Delta_B - \frac{2}{D\!-\!3} i\Delta_C
\Biggr] \; . \qquad \label{contract3}
\end{eqnarray}
Combining everything gives,
\begin{equation}
-i \widetilde{M}^2_{B2}(x;x')= - \frac{\kappa^2 H^4 a^2}{8 \pi^2} \, 
i\delta^4(x \!-\! x') \; . \label{finalMB2}
\end{equation}

\subsection{3-Point Contributions} \label{3-point}

The left hand diagram of Figure~\ref{diagrams} represent the free
expectation values of product of first variations (\ref{a}-\ref{b}) at points
$x^{\mu}$ and $x^{\prime \mu}$. Each term involves the product of a (possibly
differentiated) graviton propagator multiplied by a (possibly differentiated)
scalar propagator. The first and second derivatives of the scalar propagator are,
\begin{eqnarray}
\partial_{\mu} i\Delta(x;x') & = & -\frac{\Gamma(\frac{D}2)}{2 \pi^{\frac{D}2}}
\frac{\Delta x_{\mu}}{\Delta x^D} \; , \label{dDelta} \\
\partial_{\mu} \partial'_{\rho} i\Delta(x;x') & = & \delta^0_{\mu} \delta^0_{\rho}
\, i\delta^D(x \!-\! x') + \frac{\Gamma(\frac{D}2)}{2 \pi^{\frac{D}2}} \Biggl[
\frac{\eta_{\mu\rho}}{\Delta x^{D}} - \frac{D \Delta x_{\mu} \Delta x_{\rho}}{
\Delta x^{D+2}}\Biggr] \; . \qquad \label{ddDelta}
\end{eqnarray}
Note that the action of two time derivatives of either propagator will produce a 
delta function, which evaluates the other propagator at coincidence. We have 
already reviewed the coincidence limits (\ref{DAcoinc}-\ref{DCcoinc}) of the 
components of the graviton propagator. A key point about the scalar propagator is 
that its coincidence limit, and that of its first derivative, both vanish in 
dimensional regularization. Note also that some of the expressions in this 
subsection were checked using the symbolic manipulation program ``Cadabra'' 
\cite{Peeters:2007wn,Peeters:2018dyg}.

\subsubsection{$-i \widetilde{M}^2_{aa}(x;x')$} \label{Msubaa}

The free expectation value of the product of two factors of (\ref{a}) is,
\begin{eqnarray}
-i \widetilde{M}^2_{aa}(x;x')& \equiv & -\kappa^2 \partial_{\mu} \partial'_{\rho}
\Biggl\langle \Omega \Biggl\vert T^*\Biggl[\widehat{h}^{\mu\nu}(x) \partial_{\nu} 
\widetilde{\phi}(x) \!\times\! \widehat{h}^{\rho\sigma}(x') \partial'_{\sigma} 
\widetilde{\phi}(x') \Biggr] \Biggr\vert \Omega \Biggr\rangle \; , \qquad 
\label{Maa1} \\
& = & -\kappa^2 \partial_{\mu} \partial'_{\rho} \Biggl\{ i\Bigl[\mbox{}^{
\widehat{\mu\nu}} \Delta^{\widehat{\rho\sigma}} \Bigr](x;x') \!\times\! 
\partial_{\nu} \partial'_{\sigma} i\Delta(x;x') \Biggr\} \; . \qquad \label{Maa2}
\end{eqnarray}
Performing the inner contractions over $\nu$ and $\sigma$ gives,
\begin{eqnarray}
\lefteqn{-i \widetilde{M}^2_{aa}(x;x') = -\kappa^2 \partial_{\mu} \partial'_{\rho}
\Biggl\{ i\Delta_A \eta^{\mu\rho} \partial \!\cdot\! \partial' i\Delta + 
(i\Delta_A \!-\! i\Delta_B) \Bigl[ \delta^{\mu}_0 \delta^{\rho}_0 \nabla \!\cdot\!
\nabla' } \nonumber \\
& & \hspace{-0.5cm} + \delta^{\mu}_0 \overline{\partial}^{\rho} \partial'_0 + 
\delta^{\rho}_0 \overline{\partial}^{\prime \mu} \partial_0 + 
\overline{\eta}^{\mu\rho} \partial_0 \partial'_0\Bigr] i\Delta  - 2 \Bigl(
\frac{D \!-\!2}{D\!-\!3}\Bigr) (i\Delta_A \!-\! i\Delta_C) \delta^{\mu}_0 
\delta^{\rho}_0 \partial_0 \partial'_0 i\Delta\Biggr\} . \qquad \label{Maa3}
\end{eqnarray}
Because the scalar propagator (unlike the graviton propagator!) depends only on 
the coordinate difference $(x-x')^{\mu}$, we can convert primed to unprimed 
derivatives $\partial'_{\mu} = -\partial_{\mu}$. We can also use the propagator
equation (\ref{propeqns}) to eliminate double time derivatives,
\begin{equation}
\partial_0 \partial'_0 i\Delta(x;x') = i\delta^D(x \!-\!x') - \nabla^2 i\Delta(x;x')
\; . \label{d0sq}
\end{equation}
These reductions uncover a local part and a nonlocal one,
\begin{eqnarray}
\lefteqn{-i \widetilde{M}^2_{aa}(x;x') = \kappa^2 \partial_{\mu} \partial'_{\rho}
\Biggl\{ \Biggl[\eta^{\mu\rho} i\Delta_A - \overline{\eta}^{\mu\rho} (i\Delta_A
\!-\! i\Delta_B) + 2 \Bigl(\frac{D \!-\! 2}{D\!-\!3}\Bigr) \delta^{\mu}_0 
\delta^{\rho}_0 } \nonumber \\
& & \hspace{-0.5cm} \times (i\Delta_A \!-\! i\Delta_C)\Biggr] i\delta^D(x \!-\! x')
+ (i\Delta_A \!-\! i\Delta_B) \Bigl[ (\delta^{\mu}_0 \delta^{\rho}_0 \!+\!
\overline{\eta}^{\mu\rho} ) \nabla^2 + 2 \delta^{(\mu}_0 \overline{\partial}^{\rho)}
\partial_0 \Bigr] i\Delta \nonumber \\
& & \hspace{5.7cm} - 2 \Bigl(\frac{D \!-\!2}{D \!-\!3}\Bigr) (i\Delta_A \!-\! 
i\Delta_C) \delta^{\mu}_0 \delta^{\rho}_0 \nabla^2 i\Delta \Biggr\} . \qquad 
\label{Maa4}
\end{eqnarray}

Further reducing the local contributions of (\ref{Maa4}) is easy. The coincident 
propagators can be read off from expressions (\ref{DAcoinc}-\ref{DCcoinc}), and the 
three tensor structures can be consolidated to two using $\delta^{\mu}_0 
\delta^{\rho}_0 = \overline{\eta}^{\mu\rho} - \eta^{\mu\rho}$. The three nonlocal 
contributions of (\ref{Maa4}) require more work. We first extract derivatives using 
the relations,
\begin{eqnarray}
\partial_0 \partial_i \Bigl(\frac1{\Delta x^{D-2}} \Bigr) \!\times\! 
\frac1{\Delta x^{D-4}} & = & \frac14 \Bigl( \frac{D}{D\!-\!3}\Bigr) \partial_0 
\partial_i \Bigl( \frac1{\Delta x^{2D-6}} \Bigr) \; , \qquad \label{IDa1} \\
\nabla^2 \Bigl(\frac1{\Delta x^{D-2}} \Bigr) \!\times\! \frac1{\Delta x^{D-4}} & = &
\frac14 \Bigl[ \Bigl(\frac{D}{D\!-\!3}\Bigr) \nabla^2 -
\Bigl(\frac{D\!-\!1}{D\!-\!3}\Bigr) \partial^2\Bigr] \frac1{\Delta x^{2D-6}} \; , 
\qquad \label{IDa2}
\end{eqnarray}
We also need the identity \cite{Onemli:2002hr},
\begin{equation}
\partial^2 \frac1{\Delta x^{2D-6}} = \frac{\mu^{D-4} 4 \pi^{\frac{D}2} i
\delta^D(x \!-\!x')}{\Gamma(\frac{D}2 \!-\! 1)} - \frac12 (D\!-\!4) \partial^2
\Bigl( \frac{\ln(\mu^2 \Delta x^2)}{\Delta x^2}\Bigr) + O\Bigl( (D\!-\!4)^2\Bigr) 
. \label{IDa3}
\end{equation} 
Relations (\ref{IDa1}-\ref{IDa3}) imply,
\begin{eqnarray}
\lefteqn{(i\Delta_A \!-\! i\Delta_B) \partial_0 \partial_i i\Delta = -
\frac{H^2 \partial_0 \partial_i}{16 \pi^4} \Bigg\{ 
\frac{\frac12 \ln(\frac14 H^2 \Delta x^2) \!+\! 1}{\Delta x^2}\Biggr\} \; ,} \\
\lefteqn{(i\Delta_A \!-\! i\Delta_B) \nabla^2 i\Delta = -\frac{H^2 \nabla^2}{16 \pi^4} 
\Bigg\{ \frac{\frac12 \ln(\frac14 H^2 \Delta x^2) \!+\! 1}{\Delta x^2}\Biggr\}
+ \frac{3 H^2 \partial^2}{128 \pi^4} \Bigg\{ \frac{\ln(\mu^2 \Delta x^2)}{\Delta x^2}
\Biggr\} } \nonumber \\
& & \hspace{5.4cm} -\frac{\mu^{D-4} H^2}{16 \pi^{\frac{D}2}} \frac{(D\!-\!1) 
\Gamma(\frac{D}2)}{(D\!-\!3) (D\!-\!4)} \frac{ i\delta^D(x \!-\! x')}{
(a a')^{\frac{D}2 -2}} \; , \qquad \\
\lefteqn{(i\Delta_A \!-\! i\Delta_C) \nabla^2 i\Delta = -\frac{H^2 \nabla^2}{16 \pi^4} 
\Bigg\{ \frac{\frac12 \ln(\frac14 H^2 \Delta x^2) \!+\! 1}{\Delta x^2}\Biggr\}
+ \frac{3 H^2 \partial^2}{128 \pi^4} \Bigg\{ \frac{\ln(\mu^2 \Delta x^2)}{\Delta x^2}
\Biggr\} } \nonumber \\
& & \hspace{5.4cm} -\frac{\mu^{D-4} H^2}{16 \pi^{\frac{D}2}} \frac{2 (D\!-\!1) 
\Gamma(\frac{D}2)}{(D \!-\! 2) (D\!-\!4)} \frac{ i\delta^D(x \!-\! x')}{
(a a')^{\frac{D}2 -2}} \; , \qquad
\end{eqnarray}
where we have ignored terms that vanish at $D=4$. Putting everything together gives,
\begin{eqnarray}
\lefteqn{-i \widetilde{M}^2_{aa}(x;x') = \kappa^2 \Biggl[
\Bigl( \frac{D\!-\!1}{D\!-\!3}\Bigr) \Bigl(A_0 \!-\! \frac34 A_1\Bigr) \!+\! 
\frac{H^2}{4 \pi^2} \Biggr] \partial \!\cdot\! \partial' \Biggl\{ \frac{i
\delta^D(x \!-\! x')}{(a a')^{\frac{D}2 -2}} \Biggr\} } \nonumber \\
& & \hspace{2cm} - \kappa^2 \Biggl[ \Bigl(\frac{D\!-\!1}{D\!-\!3}
\Bigr) \Bigl(A_0 \!-\! \frac12 A_1\Bigr) \!+\! \frac{5 H^2}{16 \pi^2} \Biggr] 
\vec{\nabla} \!\cdot\! \vec{\nabla}' \Biggl\{\frac{i \delta^D(x \!-\! x')}{
(a a')^{\frac{D}2-2}} \Biggr\}  \nonumber \\
& & \hspace{-0.5cm} + \frac{\kappa^2 H^2 \partial^2 \nabla^2}{16 \pi^4} \Biggl\{ 
\frac{ \frac12 \ln(\frac14 H^2 \Delta x^2) \!+\! 1}{\Delta x^2} \Biggr\} 
\!-\! \frac{3 \kappa^2 H^2 \partial^2 (3 \partial^2 \!-\! 2 \nabla^2)}{128 \pi^4}
\Biggl\{ \frac{\ln(\mu^2 \Delta x^2)}{\Delta x^2} \Biggr\} . \qquad \label{finalMaa}
\end{eqnarray}
where the constants $A_0$ and $A_1$ are,
\begin{equation}
A_0 \equiv \frac{H^{D-2}}{(4 \pi)^{\frac{D}2}} \frac{\Gamma(D \!-\! 1)}{\Gamma(\frac{D}2)} 
\!\times\! \pi {\rm cot}\Bigl( \frac{\pi D}{2}\Bigr) \qquad , \qquad A_1 \equiv 
\frac{\mu^{D-4} H^2}{4 \pi^{\frac{D}2}} \frac{\Gamma(\frac{D}2)}{D \!-\! 4} \; . 
\label{A0A1}
\end{equation}

\subsubsection{$-i \widetilde{M}^2_{ab}(x;x')$ and 
$-i\widetilde{M}^2_{ba}(x;x')$} \label{Msubab}

Although there are two products of (\ref{a}) and (\ref{b}), we first consider the product of 
(\ref{a}) at $x^{\mu}$ with (\ref{b}) at $x^{\prime \mu}$, and then derive the crossed 
contribution by interchanging the coordinates,
\begin{eqnarray}
\lefteqn{-i \widetilde{M}^2_{ab}(x;x') } \nonumber \\
& & \hspace{-0.5cm} \equiv -\frac{\kappa^2}{4} \Bigl(\frac{D\!-\!2}{D\!-\!1}\Bigr)
\partial_{\mu} \Biggl\langle \!\!\Omega \Biggl\vert T^*\!\Biggl[\widehat{h}^{\mu\nu}(x) 
\partial_{\nu} \widetilde{\phi}(x) \!\times\! \widetilde{\phi}(x') \Bigl[
\partial'_{\rho} \partial'_{\sigma} \!\!+\! \frac{\eta_{\rho\sigma} 
\partial^{\prime 2}}{D \!-\!2} \Bigr] \widehat{h}^{\rho\sigma}(x') \Biggr] \!\Biggr\vert 
\Omega \!\!\Biggr\rangle , \qquad \label{Mab1} \\
& & \hspace{-0.5cm} = -\frac{\kappa^2}{4} \Bigl(\frac{D\!-\!2}{D\!-\!1}\Bigr) 
\partial_{\mu} \Biggl\{ \partial_{\nu} i\Delta(x;x') \!\times\! \Bigl[\partial'_{\rho} 
\partial'_{\sigma} + \frac{\eta_{\rho\sigma} \partial^{\prime 2}}{D \!-\!2} \Bigr] 
i\Bigl[\mbox{}^{\widehat{\mu\nu}} \Delta^{\widehat{\rho\sigma}} \Bigr](x;x') \Biggr\} 
, \qquad \label{Mab2} \\
& & \hspace{-0.5cm} = \frac{\kappa^2}{2} \Bigl(\frac{D\!-\!2}{D\!-\!1}\Bigr) 
\partial_{\mu} \Biggl\{ \partial_{\nu} i\Delta \Biggl[ \Bigl(\eta^{\mu\nu} 
\partial^{\prime 2} \!-\! \partial^{\prime \mu} \partial^{\prime \nu}\Bigr) 
i\Delta_A - 2 \delta^{(\mu}_0 \overline{\partial}^{\prime \nu)} \partial'_0 \Bigl[
i\Delta_A \!-\! i\Delta_B\Bigr] \nonumber \\
& & \hspace{3cm} + \frac{\delta^{\mu}_0 \delta^{\nu}_0}{D \!-\! 3} \Bigl[-(D\!-\!1)
\partial^{\prime 2} + (D \!-\!2) \nabla^{\prime 2}\Bigr] \Bigl[i\Delta_A \!-\!
i\Delta_C\Bigr] \Biggr] \Biggr\} . \qquad \label{Mab3} 
\end{eqnarray}
If we keep the $\partial_{\mu}$ derivative outside the curly brackets the terms 
inside diverge only cubicly, which reduces the number of terms that must be retained
in the expansion of the three $i\Delta_I(x;x')$. Our strategy is accordingly to
extract derivatives from the curly brackets.

The three tensor factors inside the large square brackets of (\ref{Mab3}) have the 
$3+1$ decompositions,
\begin{eqnarray}
\Bigl(\eta^{\mu\nu} \partial^{\prime 2} \!-\! \partial^{\prime \mu} \partial^{\prime \nu}
\Bigr) & = & \left( \matrix{-\nabla^{\prime 2} & \partial^{\prime}_0 \partial^{\prime}_n \cr
\partial^{\prime}_m \partial^{\prime}_0 & \delta_{mn} \partial^{\prime 2} \!-\! 
\partial^{\prime}_m \partial^{\prime}_n}\right) \; , \\
\Bigl( \delta^{\mu}_0 \overline{\partial}^{\prime \nu} \!+\! \delta^{\nu}_0 
\overline{\partial}^{\prime \mu} \Bigr) \partial'_0 & = & \left(\matrix{0 & \partial'_0 
\partial'_n \cr \partial'_m \partial'_0 & 0 }\right) \; , \\
\delta^{\mu}_0 \delta^{\nu}_0 & = & \left(\matrix{1 & 0 \cr 0 & 0}\right) \; .
\end{eqnarray}
We next use the propagator equations (\ref{propeqns}), and the fact that the coincidence
limit of $\partial_{\nu} i\Delta(x;x')$ vanishes, to contract the factor inside the large
square brackets of expression (\ref{Mab3}) into the derivative of the scalar. The results
are,
\begin{eqnarray}
\Biggl[ \qquad \Biggr]^{0\nu} \!\!\! \partial_{\nu} i\Delta \!\!\!& = &\!\!\! \Biggl[- 
\Bigl[\nabla^{\prime 2} - 2 (D\!-\!1) H^2 a^{\prime 2}\Bigr] i\Delta_C \nonumber \\
& & \hspace{-3cm} + \Bigl[\nabla^{\prime 2} - (D\!-\!2) (D\!-\!1) H a' \partial_0'\Bigr]
\Bigl( \frac{i\Delta_A \!-\!i\Delta_C}{D \!-\!3}\Bigr)\Biggr] \partial_0 i\Delta + 
\partial_0' \partial_n' i\Delta_B \partial_n i\Delta \; , \qquad \label{Mab4a} \\
\Biggl[ \qquad \Biggr]^{m\nu} \!\!\! \partial_{\nu} i\Delta \!\!\!& = &\!\!\! \partial_0' 
\partial_m' i\Delta_B \partial_0 i\Delta + \Bigl[ \delta_{mn} (D\!-\!2) H a' \partial_0' - 
\partial_m' \partial_n'\Bigr] i\Delta_A \partial_n i\Delta \; . \qquad \label{Mab4b} 
\end{eqnarray}

The next step is to act the derivatives in expressions (\ref{Mab4a}-\ref{Mab4b}),
combine terms and extract derivatives. These manipulations are very tedious so we only
present the results, dropping terms that vanish for $D=4$,
\begin{eqnarray}
\lefteqn{ \Biggl[ \qquad \Biggr]^{0\nu} \!\!\! \partial_{\nu} i\Delta =
\frac{\Gamma^2(\frac{D}2) \partial_0}{8 \pi^D} \Biggl\{ \frac{\partial^{\prime 2}}{2
(D\!-\!2)^2} \Biggl[\frac1{(a a' \Delta x^4)^{\frac{D}2 - 1}}\Biggr] \!+\!
\frac{(D^2 \!+\! 10 D \!-\! 8) H^2 a^{\prime 2}}{8 (D\!-\!2)^2 (a a' \Delta x^4)^{\frac{D}2-1}} 
} \nonumber \\
& & \hspace{1.5cm} - \frac12 \Bigl( \frac{3D\!-\!4}{D\!-\!1}\Bigr) 
\frac{H a' \Delta x_0}{(a a')^{\frac{D}2 -1} \Delta x^{2D-2}} + \frac12 \Bigl(\frac{D\!-\!2}{D\!-\!1}
\Bigr) \frac{H^2 \Delta x_0^2}{(a a')^{\frac{D}2-2} \Delta x^{2D-2}} \nonumber \\
& & \hspace{1cm} - \frac{(D\!+\!1) (D^2 \!-\! 7D \!+\! 8) H^2}{8 (D\!-\!1) (D\!-\! 2) 
(a a')^{\frac{D}2 -2} \Delta x^{2D-4}} - \Bigl( \frac{D\!-\!1}{D \!-\! 2}\Bigr) 
\frac{H^3 a' \Delta x_0}{(a a')^{\frac{D}2 -2} \Delta x^{2D-4}} \Biggr\} \nonumber \\
& & \hspace{-0.5cm} + \frac{\Gamma^2(\frac{D}2)}{8 \pi^D} \Biggl\{
\frac12 \Bigl( \frac{D\!-\!2}{D\!-\!1}\Bigr) \frac{[(D\!-\!1) H a \!+\! (2D\!-\!3) H a'
\!-\! \frac12 (D\!-\!2) H^2 a a' \Delta x_0]}{(a a')^{\frac{D}2 -1} \Delta x^{2D-2}}
\nonumber \\
& & \hspace{1.5cm} + \frac{(D\!-\!4) H^2}{4 (a a')^{\frac{D}2 -2}} \Biggl[ \Bigl(
\frac{D\!-\!2}{D\!-\!1}\Bigr) \frac{H a \Delta x_0^2}{\Delta x^{2D -2}} + 
\frac{(D^2 \!-\! 7 D \!+\! 8) (D\!+\! 1)}{4 (D\!-\!2) (D\!-\!1)} \frac{Ha}{\Delta x^{2D-4}}
\nonumber \\
& & \hspace{7.5cm} - 2 \Bigl(\frac{D\!-\!1}{D\!-\!2}\Bigr) \frac{H^2 a a' \Delta x_0}{
\Delta x^{2D-4}} \Biggr] \Biggr\} \; , \qquad \label{Mab4a2} \\
\lefteqn{ \Biggl[ \qquad \Biggr]^{m\nu} \!\!\! \partial_{\nu} i\Delta =
\frac{\Gamma^2(\frac{D}2) \partial_m}{8 \pi^D} \Biggl\{\frac{-\partial^{\prime 2}}{2
(D\!-\!2)^2} \Biggl[\frac1{(a a' \Delta x^4)^{\frac{D}2 - 1}}\Biggr] \!-\! 
\frac{(5D \!-\!12) H^2 {a'}^2}{{8 (D\!-\!2) (a a' \Delta x^4)}^{\frac{D}2 -1}} } 
\nonumber \\
& & \hspace{1.5cm} + \frac12 \Bigl( \frac{3D\!-\!4}{D\!-\!1}\Bigr) 
\frac{H a' \Delta x_0}{(a a')^{\frac{D}2 -1} \Delta x^{2D-2}} - \frac12 \Bigl(
\frac{D\!-\!2}{D\!-\!1}\Bigr) \frac{H^2 \Delta x_0^2}{(a a')^{\frac{D}2-2} 
\Delta x^{2D-2}} \nonumber \\
& & \hspace{2cm} + \frac{D H^2 [-(D\!-\!3) + (D\!-\!2) H a' \Delta x_0]}{8 (D\!-\!2)
(a a')^{\frac{D}2 -2} \Delta x^{2D-4}} \Biggr\} . \qquad \label{Mab4b2}
\end{eqnarray}

We now substitute expressions (\ref{Mab4a2}-\ref{Mab4b2}) in (\ref{Mab3}) and 
extract derivatives from each term until the expression becomes integrable. At that
point divergences are localized and the unregulated limit is taken on the nonlocal
terms. Relevant identities are,
\begin{eqnarray}
\frac1{\Delta x^{2D-2}} & \longrightarrow & \frac{K \partial^2 i \delta^D(x\!-\!x')}{
2 (D\!-\!2)^2} - \frac{\partial^4}{32} \Biggl[ \frac{ \ln(\mu^2 \Delta x^2)}{
\Delta x^2} \Biggr] , \label{EXD1} \\
\frac{\Delta x_0}{\Delta x^{2 D-2}} & \longrightarrow & -\frac{K \partial_0 i
\delta^D(x \!-\!x')}{2 (D\!-\!2)} + \frac{\partial_0 \partial^2}{16} \Biggl[ 
\frac{ \ln(\mu^2 \Delta x^2)}{\Delta x^2} \Biggr] , \label{EXD2} \\
\frac{\Delta x_0^2}{\Delta x^{2 D-2}} & \longrightarrow & -
\frac{K i\delta^D(x \!-\!x')}{2 (D\!-\!2)} + \frac{\partial^2}{16} \Biggl[ 
\frac{ \ln(\mu^2 \Delta x^2) \!-\! 2}{\Delta x^2} \Biggr] + \frac{\nabla^2}{8}
\Biggl[ \frac1{\Delta x^2}\Biggr] , \label{EXD3} \\
\frac1{\Delta x^{2D-4}} & \longrightarrow & K i\delta^D(x \!-\! x') -
\frac{\partial^2}{4} \Biggl[ \frac{ \ln(\mu^2 \Delta x^2)}{\Delta x^2} \Biggr] \; , 
\label{EXD4} \\
\frac{\Delta x_0}{\Delta x^{2D -4}} & \longrightarrow & -\frac{\partial_0}{2}
\Biggl[ \frac1{\Delta x^2} \Biggr] , \label{EXD5}
\end{eqnarray}
where the constant $K$ is,
\begin{equation}
K \equiv \frac{\mu^{D-4} 4 \pi^{\frac{D}2}}{2 (D\!-\!3) (D\!-\!4) \Gamma(\frac{D}2 -1)}
\; . \label{Kdef}
\end{equation}
The final result is,
\begin{eqnarray}
\lefteqn{-i \widetilde{M}^2_{ab}(x;x') = \frac{-\kappa^2 A_1}{(D\!-\!1) (D\!-\!3)} 
\Biggl\{ \frac{\partial^2 \partial^{\prime 2}}{8 H^2} \Biggl[ 
\frac{i \delta^D(x \!-\! x')}{(a a')^{\frac{D}2 -1}} \Biggr] + \frac{3 \partial^2}{D\!-\!1} 
\Biggl[\frac{i \delta^D(x \!-\!x')}{(a a')^{\frac{D}2 -2}} \Biggr] \Biggr\} } \nonumber \\
& & \hspace{3cm} + \frac{\kappa^2 H^2}{24 \pi^2} \Biggl\{ -\frac{11}{6} \nabla^2 -
\frac{17}{12} \partial^2 - \frac12 H a' \partial_0\Biggr\} i\delta^4(x \!-\! x') 
\nonumber \\
& & \hspace{1.5cm} + \frac{\kappa^2 H^2}{96 \pi^4} \Biggl\{ \frac{\partial^2 
{\partial'}^2}{8 H^2} \Biggl( \frac1{a a'} \partial^2 \Biggl[ \frac{\ln(\mu^2 
\Delta x^2)}{\Delta x^2}\Biggr] \Biggr) + \partial^2 {\partial'}^2 \Biggl[
\frac{\ln(\mu^2 \Delta x^2)}{\Delta x^2}\Biggr] \Biggr\} . \qquad \label{Mab5}
\end{eqnarray}

The result for $-i \widetilde{M}^2_{ba}(x;x')$ follows by simply interchanging $x^{\mu}$
and ${x'}^{\mu}$. Most of the terms in (\ref{Mab5}) are already reflection symmetric, and
a few simple reductions suffice for those which are not,
\begin{eqnarray}
\frac{3 A_1 (\partial^2 \!+\! {\partial'}^2)}{(D\!-\!1)^2 (D\!-\!3)} \Biggl[\frac{i 
\delta^D(x \!-\!x')}{(a a')^{\frac{D}2 -2}} \Biggr] & \longrightarrow & -
\frac{6 A_1 \partial \!\cdot\! \partial'}{(D\!-\!1)^2 (D\!-\!3)} \Biggl[\frac{i 
\delta^D(x \!-\!x')}{(a a')^{\frac{D}2 -2}} \Biggr] \nonumber \\
& & \hspace{2.8cm} + \frac{H^4 a^2 i\delta^4(x \!-\! x')}{12 \pi^2} \; , \qquad \\
-\frac12 H (a' \partial_0 \!+\! a \partial'_0) i\delta^4(x \!-\! x') & = & -\frac12 
H^2 a^2 i \delta^4(x \!-\! x') \; . \qquad 
\end{eqnarray}
The final result for sum is,
\begin{eqnarray}
\lefteqn{-i \widetilde{M}^2_{ab+ba} = \frac{\kappa^2 A_1}{(D\!-\!1) (D\!-\!3)} 
\Biggl\{ -\frac{\partial^2 \partial^{\prime 2}}{4 H^2} \Biggl[ 
\frac{i \delta^D(x \!-\! x')}{(a a')^{\frac{D}2 -1}} \Biggr] + \frac{6 \partial 
\!\cdot\! \partial'}{D\!-\!1} \Biggl[\frac{i \delta^D(x \!-\!x')}{(a a')^{\frac{D}2 -2}} 
\Biggr] \Biggr\} } \nonumber \\
& & \hspace{3cm} + \frac{\kappa^2 H^2}{24 \pi^2} \Biggl\{ -\frac{11}{3} \nabla^2 -
\frac{17}{6} \partial^2 - \frac52 H^2 a^2 \Biggr\} i\delta^4(x \!-\! x') \nonumber \\
& & \hspace{1.2cm} + \frac{\kappa^2 H^2}{96 \pi^4} \Biggl\{ \frac{\partial^2 
{\partial'}^2}{4 H^2} \Biggl( \frac1{a a'} \partial^2 \Biggl[ \frac{\ln(\mu^2 
\Delta x^2)}{\Delta x^2}\Biggr] \Biggr) + 2 \partial^2 {\partial'}^2 \Biggl[
\frac{\ln(\mu^2 \Delta x^2)}{\Delta x^2}\Biggr] \Biggr\} . \qquad \label{finalMab}
\end{eqnarray}

\subsubsection{$-i \widetilde{M}^2_{bb}(x;x')$} \label{Msubbb}

The last of the 3-point contributions is the free expectation value of the product 
of two (\ref{b}) interactions. We begin by reducing this to propagators and performing
the initial contractions, 
\begin{eqnarray}
\lefteqn{-i \widetilde{M}^2_{bb}(x;x') \equiv -\frac{\kappa^2}{16} 
\Bigl(\frac{D\!-\!2}{D\!-\!1}\Bigr)^2 } \nonumber \\
& & \hspace{-0.2cm} \times \!\Biggl\langle \!\Omega \Biggl\vert T^*\!\Biggl[\widetilde{\phi}(x)
\Bigl[\partial_{\mu} \partial_{\nu} + \frac{\eta_{\mu\nu} \partial^2}{D\!-\!2}\Bigr]
\widehat{h}^{\mu\nu}(x) \!\times\! \widetilde{\phi}(x') \Bigl[\partial'_{\rho} 
\partial'_{\sigma} \!+\! \frac{\eta_{\rho\sigma} \partial^{\prime 2}}{D \!-\!2} \Bigr] 
\widehat{h}^{\rho\sigma}(x') \Biggr] \Biggr\vert \Omega \!\Biggr\rangle , \qquad 
\label{Mbb1} \\
& & \hspace{-0.5cm} = -\frac{\kappa^2}{16} \Bigl(\frac{D\!-\!2}{D\!-\!1}\Bigr)^2 
i\Delta(x;x') \Bigl[\partial_{\mu} \partial_{\nu} \!+\! \frac{\eta_{\mu\nu} \partial^2
}{D\!-\!2}\Bigr] \Bigl[\partial'_{\rho} \partial'_{\sigma} \!+\! \frac{\eta_{\rho\sigma} 
\partial^{\prime 2}}{D \!-\!2} \Bigr] i\Bigl[\mbox{}^{\widehat{\mu\nu}} 
\Delta^{\widehat{\rho\sigma}} \Bigr](x;x') , \qquad \label{Mbb2} \\
& & \hspace{-0.5cm} = -\frac{\kappa^2}{16} \Bigl(\frac{D\!-\!2}{D\!-\!1}\Bigr)^2 
i\Delta \Biggl\{ \Biggl[ 2 (\partial \!\cdot\! \partial')^2 - 2 \Bigl(
\frac{2D \!-\! 3}{D \!-\!2}\Bigr) \partial^2 {\partial'}^2 \Biggr] i\Delta_A 
\nonumber \\
& & \hspace{.5cm} + 4 \partial_0 \partial_0' \vec{\nabla} \!\cdot\! \vec{\nabla}' 
(i\Delta_A \!-\! i\Delta_B) - \frac{2}{D\!-\!3} \Biggl[ \frac{(D\!-\!1)^2}{(D\!-\!2)} 
\partial^2 {\partial'}^2 - (D\!-\!1) \partial^2 {\nabla'}^2 \nonumber \\
& & \hspace{3.5cm} - (D\!-\!1) {\partial'}^2 \nabla^2
+ (D\!-\!2) \nabla^2 {\nabla'}^2 \Biggr] (i\Delta_A \!-\! i\Delta_C)\Biggr\} . \qquad
\label{Mbb3}
\end{eqnarray}
Because this contribution begins with no derivatives extracted, it is necessary to
retain up to second order terms in the propagator expansions (\ref{propA}-\ref{propC}).

At this stage our strategy is to eliminate factors of $\partial^2$ and ${\partial'}^2$
using the propagator equations (\ref{propeqns}). We can also eliminate factors of 
$\partial_0 \partial'_0$ by employing the reflection identities \cite{Tsamis:1992zt,
Miao:2009hb,Miao:2010vs},
\begin{eqnarray}
\partial_0 i\Delta_A(x;x') &=& -\Bigl[\partial'_0 + (D\!-\!2) H a'\Bigr]
i\Delta_B(x;x') \; , \label{refl1} \\
(\partial_0 \!+\! H a) i\Delta_B(x;x') &=& -\Bigl[ \partial_0' + (D\!-\!3) H a'\Bigr] 
i\Delta_C(x;x') \; . \label{refl2}
\end{eqnarray}
Significant reductions are,
\begin{eqnarray}
\partial_0 \partial'_0 i\Delta_A & = & \frac{i\delta^D(x \!-\!x')}{a^{D-2}} - \nabla^2
i\Delta_B \; , \\
\partial_0 \partial'_0 i\Delta_B & = & \frac{i\delta^D(x \!-\!x')}{a^{D-2}} - \nabla^2
i\Delta_A + (D\!-\!2) H \Bigl[a \partial_0 i\Delta_A \!-\! a' \partial_0 i\Delta_B \Bigr]
\; , \qquad \\ 
\partial_0 \partial'_0 i\Delta_C & = & \frac{i\delta^D(x \!-\!x')}{a^{D-2}} - \nabla^2
i\Delta_B + (D\!-\!3) H \Bigl[a \partial_0 i\Delta_B \!-\! a' \partial_0 i\Delta_C\Bigr]
\nonumber \\
& & \hspace{6.5cm} + (D\!-\!3) H^2 a^2 i\Delta_B \; . \qquad 
\end{eqnarray}
The reflection identities (\ref{refl1}-\ref{refl2}) can also be used to reach a 
manifestly reflection invariant form. Note also that we can ignore factors of 
$i\delta^D(x - x')$ because the coincidence limit of $i\Delta(x;x')$ vanishes in 
dimensional regularization. The final result for this stage of the reduction is,
\begin{eqnarray}
\lefteqn{-i \widetilde{M}^2_{bb}(x;x') \equiv -\frac{\kappa^2}{16} 
\Bigl(\frac{D\!-\!2}{D\!-\!1}\Bigr)^2 i\Delta \Biggl\{ -2(D\!-\!2) H (a \partial'_0 \!+\!
a' \partial_0) \nabla^2 i\Delta_B } \nonumber \\
& & \hspace{-0.5cm} + 2 (D\!-\!2) (D\!-\!1) H^2 a a' \nabla^2 i\Delta_B -2(D\!-\!1)^2 
(D\!-\!4) H^3 a a' (a \partial'_0 \!+\! a' \partial_0) i\Delta_C \nonumber \\
& & \hspace{-0.5cm} - \frac{2 D (D\!-\!1)^2 (D\!-\!3) (D\!-\! 4)}{(D\!-\!2)} H^4 a^2 
{a'}^2 i\Delta_C + \frac{2 (D\!-\!1) (D\!-\!2)}{(D\!-\!3)} H (a \partial_0 \!+\! a' 
\partial'_0)  \nonumber \\
& & \hspace{1cm} \times \nabla^2 (i\Delta_A \!-\! i\Delta_C) -4 (D\!-\!1) H^2 a a' 
(2 \!+\! H^2 a a' \Delta x_0^2) \nabla^2 i\Delta_C \nonumber \\
& & \hspace{7cm} - 2 \Bigl( \frac{D\!-\!2}{D\!-\!3}\Bigr) \nabla^4 (i \Delta_A \!-\! 
i\Delta_C) \; . \qquad \label{Mbb4} 
\end{eqnarray}

The next step is to act the derivatives using the identities,
\begin{eqnarray}
(a \partial'_0 \!+\! a' \partial_0) \Biggl[ \frac1{(a a' \Delta x^2)^N} \Biggr]
& = & -\frac{2 N H \Vert \Delta \vec{x}\Vert^2}{(a a')^{N-1} \Delta x^{2N+2}} \; , 
\label{DID1} \\
\nabla^2 \Biggl[ \frac1{\Delta x^{2N}} \Biggr] & = & -\frac{2 N (D\!-\!1)}{
\Delta x^{2N+2}} + \frac{4 N (N \!+\! 1) \Vert \Delta \vec{x}\Vert^2}{\Delta x^{2N+4}} 
\; , \qquad \label{DID2} \\
\nabla^2 \Biggl[ \frac{\Vert \Delta \vec{x}\Vert^2}{\Delta x^{2N+2}} \Biggr] & = &
\frac{2 (D\!-\!1)}{\Delta x^{2N+2}} - \frac{2 (N \!+\! 1) (D \!+\! 3) \Vert \Delta 
\vec{x}\Vert^2}{\Delta x^{2N+4}} \nonumber \\
& & \hspace{2.8cm} + \frac{4 (N \!+\! 1) (N \!+\! 2) \Vert \Delta \vec{x} \Vert^4}{
\Delta x^{2N+6}} \; . \qquad \label{DID3}
\end{eqnarray}
The final step is to extract derivatives using expressions (\ref{EXD1}-\ref{EXD5})
and the new relations,
\begin{eqnarray}
\frac{\Vert \Delta \vec{x}\Vert^2}{\Delta x^{2D}} \!\!& \longrightarrow &\!\! 
\Biggl[\frac{\nabla^2}{4 (D\!-\!2) (D\!-\!1)} + \frac{\partial^2}{4 (D\!-\!2)^2} 
\Biggr] K i \delta^D(x\!-\!x') \nonumber \\
& & \hspace{4.8cm} - \Biggl[ \frac{\nabla^2 \partial^2}{96} \!+\! 
\frac{\partial^4}{64}\Biggr] \Biggl[ \frac{ \ln(\mu^2 \Delta x^2)}{\Delta x^2} 
\Biggr] , \qquad \label{SD1} \\
\frac{\Vert \Delta \vec{x}\Vert^4}{\Delta x^{2D+2}} \!\!& \longrightarrow \!\!& 
\Biggl[\frac{(D \!+\!1) \nabla^2}{4 (D\!-\!2) (D\!-\!1) D} + \frac{(D\!+\!1) 
\partial^2}{8 (D\!-\!2)^2 D} \Biggr] K i \delta^D(x\!-\!x') \nonumber \\
& & \hspace{2cm} - \Biggl[ \frac{5 \nabla^2 \partial^2}{384} \!+\! 
\frac{5 \partial^4}{512}\Biggr] \Biggl[ \frac{ \ln(\mu^2 \Delta x^2)}{\Delta x^2} 
\Biggr] + \frac{\nabla^4}{384} \Biggl[ \frac1{\Delta x^2} \Biggr] , \qquad 
\label{SD2} \\
\frac{\Vert \Delta \vec{x}\Vert^2}{\Delta x^{2D-2}} \!\!& \longrightarrow &\!\! 
\frac12 \Bigl( \frac{D\!-\!1}{D\!-\!2}\Bigr) K i \delta^D(x \!-\! x') - 
\frac{3 \partial^2}{16} \Biggl[ \frac{ \ln(\mu^2 \Delta x^2)}{\Delta x^2} \Biggr]
+ \frac{\nabla^2}{8} \Biggl[ \frac1{\Delta x^2} \Biggr] , \qquad \label{SD3} \\
\frac{\Vert \Delta \vec{x}\Vert^4}{\Delta x^{2D}} \!\!& \longrightarrow &\!\! 
\frac14 \Bigl( \frac{D\!+\!1}{D\!-\!2}\Bigr) K i \delta^D(x \!-\! x') \nonumber \\
& & \hspace{0.5cm} - \frac{5 \partial^2}{32} \Biggl[ \frac{ \ln(\mu^2 \Delta x^2)}{
\Delta x^2} \Biggr] + \frac{5 \nabla^2}{24} \Biggl[ \frac1{\Delta x^2} \Biggr] 
- \frac{\nabla^4}{96} \Bigl[ \ln(\mu^2 \Delta x^2)\Bigr] , \qquad \label{SD4} 
\end{eqnarray}
where we recall the definition (\ref{Kdef}) of the constant $K$. Amazingly, the
final result turns out to be both finite and local,
\begin{equation}
-i \widetilde{M}^2_{bb}(x;x') = \frac{\kappa^2 H^2}{32 \pi^2} \Biggl\{
\frac13 \partial^2 - \frac49 \nabla^2 + \frac23 H^2 a^2\Biggr\} 
i \delta^4(x \!-\! x') \; . \label{finalMbb}
\end{equation}

\subsection{Total Primitive Contribution} \label{total}

The total primitive contribution comes from combining expressions 
(\ref{finalMA}), (\ref{finalMB1}), (\ref{finalMB2}), (\ref{finalMaa}), 
(\ref{finalMab}) and (\ref{finalMbb}),
\begin{eqnarray}
\lefteqn{-i \widetilde{M}^2_{\rm prim}(x;x') = \frac{-\kappa^2 A_1 \partial^2 
\partial^{\prime 2}}{4 (D\!-\!1) (D\!-\!3) H^2} \Biggl[ \frac{i \delta^D(x \!-\! x')
}{(a a')^{\frac{D}2 -1}} \Biggr] \!-\! \frac{19 \kappa^2 A_1}{12} \partial \!\cdot\!
\partial' \Biggl[ \frac{i \delta^D(x \!-\! x')}{(a a')^{\frac{D}2 -2}} \Biggr] } 
\nonumber \\
& & \hspace{-0.7cm} + \kappa^2 \! \Bigl( A_0 \!+\! \frac32 A_1\Bigr) \vec{\nabla} 
\!\cdot\! \vec{\nabla}' \Biggl[ \frac{i \delta^D(x \!-\! x')}{(a a')^{\frac{D}2 -2}} 
\Biggr] \!+\! \frac{\kappa^2 H^2}{12 \pi^2} \Biggl[ \frac{139}{24} \partial^2 \!-\! 
\frac{17}{4} \nabla^2 \!+\! \frac12 H^2 a^2 \Biggr] i \delta^4(x \!-\! x') \nonumber \\
& & \hspace{-0.5cm} + \frac{\kappa^2 \partial^2 {\partial'}^2}{384 \pi^4} \Biggl(
\frac1{a a'} \partial^2 \Biggl[ \frac{\ln(\mu^2 \Delta x^2)}{\Delta x^2}\Biggr]
\Biggr) - \frac{\kappa^2 H^2 (19 \partial^4 \!-\! 18 \nabla^2 \partial^2)}{384 \pi^4}
\Biggl[ \frac{\ln(\mu^2 \Delta x^2)}{\Delta x^2}\Biggr] \nonumber \\
& & \hspace{4cm} + \frac{\kappa^2 H^2 \partial^2 \nabla^2}{16 \pi^4} \Biggl[
\frac{\frac12 \ln(\frac14 H^2 \Delta x^2) \!+\! 1}{\Delta x^2}\Biggr] \ . \qquad 
\label{Mprim}
\end{eqnarray}
Recall that the constants $A_0$ and $A_1$ were defined in expression (\ref{A0A1}).
Note also that we have expanded the complications functions of $D$ that multiply 
$A_0$ and $A_1$ around $D=4$ and used the limits,
\begin{equation}
\lim_{D \rightarrow 4} (D-4) \!\times\! A_0 = \lim_{D \rightarrow 4} (D-4) 
\!\times\! A_1 = \frac{H^2}{4 \pi^2} \; .
\end{equation}
Finally, note that the flat space limit of our result comes entirely from the 
mixed contribution (\ref{finalMab}),
\begin{equation}
\lim_{H \rightarrow 0} \Bigl[ -i \widetilde{M}^2_{\rm prim}(x;x')\Bigr] = -
\frac{\kappa^2 \mu^{D-4} \Gamma(\frac{D}2) \partial^4 \, i\delta^D(x\!-\!x')
}{16 \pi^{\frac{D}2} (D\!-\!1) (D\!-\!3) (D\!-\!4)}  + 
\frac{\kappa^2 \partial^6}{384 \pi^4} \Biggl[ \frac{\ln(\mu^2 \Delta x^2)}{
\Delta x^2}\Biggr] \; . \label{flatlim}
\end{equation}
It is cheering to see that the finite part of this obeys the flat space 
correspondence limit found by Fr\"ob, cf. eqn (58) of \cite{Frob:2017apy}.

\section{Renormalization} \label{renorm}

On de Sitter background, and using our gauge \cite{Tsamis:1992xa,Woodard:2004ut}, 
four counterterms are required to renormalize $-i M^2(x;x')$ at one loop order.
Each involves two scalars and four derivatives distributed variously over the
scalars and metrics. How to express them is motivated by the effects of conformal
rescaling (\ref{conformal}),
\begin{eqnarray}
\sqrt{-g} R & = & a^{D-2} \Biggl[\sqrt{-\widetilde{g}} \widetilde{R} - 
2 (D \!-\! 1) H a \partial_{\mu} \Bigl( \sqrt{-\widetilde{g}} \, \widetilde{g}^{\mu 0} 
\Bigr) \nonumber \\
& & \hspace{3.5cm} - D (D \!-\! 1) H^2 a^2 \sqrt{-\widetilde{g}} \, \widetilde{g}^{00} 
\Biggr] \; , \qquad \\
\square \phi - \frac14 \Bigl( \frac{D\!-\!2}{D\!-\!1}\Bigr) R \phi & = & 
\frac1{a^{\frac{D}2+1}} \Biggl[ \widetilde{\square} \widetilde{\phi} - \frac14 \Bigl(
\frac{D\!-\!2}{D\!-\!1}\Bigr) \widetilde{R} \widetilde{\phi}\Biggr] \; , \qquad
\end{eqnarray}
where the covariant scalar d'Alembertian is,
\begin{equation}
\square \equiv \frac{1}{\sqrt{-g}} \, \partial_{\mu} \Bigl( \sqrt{-g} \, 
g^{\mu\nu} \partial_{\nu} \Bigr) \; . \label{square}
\end{equation}
The best arrangement of counterterms seems to be,
\begin{eqnarray}
\lefteqn{\Delta \mathcal{L} = -\frac{\alpha}{2} \Biggl[ \square \phi \!-\! \frac14
\Bigl(\frac{D\!-\!2}{D\!-\!1}\Bigr) R \phi \Biggr]^2 \sqrt{-g} -\frac{\beta}{2} 
\Biggl[ \square \phi - \frac14 \Bigl( \frac{D\!-\!2}{D\!-\!1}\Bigr) R \phi\Biggr]
\frac{\phi R \sqrt{-g}}{D (D \!-\! 1)} } \nonumber \\
& & \hspace{5cm} -\frac{\gamma}{2} \partial_i \phi \partial_j \phi g^{ij} 
\frac{R \sqrt{-g}}{D (D \!-\!1)} - \frac{\delta}{2} \frac{\phi^2 R^2 \sqrt{-g}}{
D^2 (D \!-\! 1)^2} \; . \qquad \label{cterms}
\end{eqnarray}
The noncovariant term proportional to $\gamma$ is the price of using a de Sitter
breaking gauge. Note also that a general metric background might require additional 
counterterms involving other curvatures which degenerate to the Ricci scalar on de
Sitter, $R_{\mu\nu\rho\sigma} = \frac1{D (D-1)} (g_{\mu\rho} g_{\nu\sigma} -
g_{\mu\sigma} g_{\nu\rho}) R$. 

Specializing the counter-Lagrangian (\ref{cterms}) to de Sitter gives,
\begin{equation}
\Delta \mathcal{L}\Bigl\vert_{\rm de\ Sitter} = -\frac{\alpha (\partial^2 
\widetilde{\phi})^2}{2 a^2} - \frac{\beta}{2} \partial^2 \widetilde{\phi} H^2
\widetilde{\phi} - \frac{\gamma}{2} \partial_i \widetilde{\phi} \partial_i
\widetilde{\phi} H^2 - \frac{\delta}{2} \widetilde{\phi}^2 H^4 a^2 \; . 
\label{ctermsdS}
\end{equation}
The second variations of each of the four terms in the counter-action gives,
\begin{eqnarray}
\frac{i \delta^2 \Delta S_{\alpha}}{\delta \widetilde{\phi}(x) \delta 
\widetilde{\phi}(x')} & = & -\alpha \partial^2 \partial^{\prime 2} \Bigl[
\frac{i \delta^D(x \!-\! x')}{a a'} \Bigr] \; , \label{ctalpha} \\
\frac{i \delta^2 \Delta S_{\beta}}{\delta \widetilde{\phi}(x) \delta 
\widetilde{\phi}(x')} & = & -\beta H^2 \partial^2 i \delta^D(x \!-\! x') 
\; , \label{ctbeta} \\
\frac{i \delta^2 \Delta S_{\gamma}}{\delta \widetilde{\phi}(x) \delta 
\widetilde{\phi}(x')} & = & +\gamma H^2 \nabla^2 i \delta^D(x \!-\! x')
\; , \label{ctgamma} \\
\frac{i \delta^2 \Delta S_{\delta}}{\delta \widetilde{\phi}(x) \delta 
\widetilde{\phi}(x')} & = & -\delta H^4 a^2 i \delta^D(x \!-\! x') \; . 
\label{ctdelta}
\end{eqnarray}
Comparison with the primitive result (\ref{Mprim}) implies the following
values for the four counterterms,
\begin{eqnarray}
\alpha & = & \frac{-\kappa^2 A_1}{4 (D\!-\!1) (D\!-\! 3) H^2} + \alpha_{\rm fin}
\; , \qquad \label{alpha} \\
\beta & = & \frac{19 \kappa^2 A_1}{12 H^2} + \frac{139 \kappa^2}{288 \pi^2} +
\beta_{\rm fin} \; , \qquad \label{beta} \\
\gamma & = & \frac{\kappa^2}{H^2} \Bigl( A_0 \!+\! \frac32 A_1\Bigr) +
\frac{17 \kappa^2}{48 \pi^2} + \gamma_{\rm fin} \; , \qquad \label{gamma} \\
\delta & = & 0 + \frac{\kappa^2}{24 \pi^2} + \delta_{\rm fin} \; . \qquad
\label{delta}
\end{eqnarray}
With these choices the renormalized result becomes,
\begin{eqnarray}
\lefteqn{-i \widetilde{M}^2_{\rm ren}(x;x') = \kappa^2 \partial^2 \partial^{\prime 2}
\Biggl\{ \Biggl[ \frac{\ln(a a')}{96 \pi^2} \!-\! \alpha_{\rm fin}\Biggr] 
\frac{i \delta^4(x \!-\! x')}{a a'} \Biggr\} } \nonumber \\
& & \hspace{-0.5cm} + \kappa^2 H^2 \partial \!\cdot\! \partial' \Biggl\{ \Biggl[ 
\frac{19 \ln(a a')}{96 \pi^2} \!+\! \beta_{\rm fin}\Biggr] i \delta^4(x \!-\! x') 
\Biggr\} \nonumber \\
& & \hspace{-0.5cm} - \kappa^2 H^2 \vec{\nabla} \!\cdot\! \vec{\nabla}' \Biggl\{ 
\Biggl[ \frac{5 \ln(a a')}{16 \pi^2} \!+\! \gamma_{\rm fin}\Biggr] i \delta^4(x \!-\! x') 
\Biggr\} - \delta_{\rm fin} \kappa^2 H^4 a^2 i\delta^4(x \!-\! x') \nonumber \\
& & \hspace{-0.5cm} + \frac{\kappa^2 \partial^2 {\partial'}^2}{384 \pi^4} \Biggl(
\frac1{a a'} \partial^2 \Biggl[ \frac{\ln(\mu^2 \Delta x^2)}{\Delta x^2}\Biggr]
\Biggr) - \frac{\kappa^2 H^2 (19 \partial^4 \!-\! 18 \nabla^2 \partial^2)}{384 \pi^4}
\Biggl[ \frac{\ln(\mu^2 \Delta x^2)}{\Delta x^2}\Biggr] \nonumber \\
& & \hspace{6cm} + \frac{\kappa^2 H^2 \partial^2 \nabla^2}{16 \pi^4} \Biggl[
\frac{\frac12 \ln(\frac14 H^2 \Delta x^2) \!+\! 1}{\Delta x^2}\Biggr] \ . \qquad 
\label{Mren}
\end{eqnarray}

\section{Conclusions} \label{conclude}

The point of this exercise has been to begin the process of purging gauge dependence
from the linearized effective field equations in cosmology by including quantum
gravitational corrections from the source which disturbs the effective field and
from the observer who measures the disturbance \cite{Miao:2017feh}. To simplify the
tensor algebra it makes sense to work with the effective field equations for a scalar.
We might have employed the existing result for the self-mass of a massless, minimally 
coupled scalar \cite{Kahya:2007bc}. However, that is known to cause no secular growth 
for the scalar mode function \cite{Kahya:2007cm}, and the classical solution for the
exchange potential \cite{Glavan:2019yfc} is so complicated that computing its one
loop correction would be daunting. The next simplest sort of scalar is the conformally
coupled case; with an arbitrary $R \phi^2$ coupling the scalar propagator becomes much
more complicated. 

Our result for the one graviton loop correction to the self-mass of a conformally 
coupled scalar is equation (\ref{Mren}). The linearized, effective field equation
for this scalar is,
\begin{equation}
\partial^2 \widetilde{\phi}(x) - \int \!\! d^4x' \, \widetilde{M}^2(x;x') 
\widetilde{\phi}(x') = \widetilde{J}(x) \; , \label{EFE}
\end{equation}
where tildes denote conformal re-scaling (\ref{conformal}) and we employ the 
Schwinger-Keldysh formalism \cite{Schwinger:1960qe,Mahanthappa:1962ex,Bakshi:1962dv,
Bakshi:1963bn,Keldysh:1964ud,Chou:1984es,Jordan:1986ug,Calzetta:1986ey} to make the 
effective field equation both real and causal. This is a diagrammatic technique for
computing true expectation values which is almost as simple as the Feynman rules  
that produce the sorts of in-out matrix elements we have computed in this paper. For 
our purposes the rules are \cite{Ford:2004wc}:
\begin{itemize}
\item{Every line carries a $\pm$ polarity corresponding to the usual case of a field 
being evolved forward in time ($+$) or being evolved backwards ($-$);}
\item{The $++$ propagator agrees with the Feynman propagator, and the $--$ propagator
is its complex conjugate, while the $+-$ and $-+$ propagators are obtained by replacing
the interval $\Delta x^2$ in expression (\ref{propphi}) with,
\begin{eqnarray}
\Delta x^2_{+-} & \equiv & \Bigl\Vert \vec{x} \!-\! \vec{x}'\Bigr\Vert^2 - \Bigl(\eta
\!-\! \eta' \!+\! i \epsilon\Bigr)^2 \; , \\
\Delta x^2_{-+} & \equiv & \Bigl\Vert \vec{x} \!-\! \vec{x}'\Bigr\Vert^2 - \Bigl(\eta
\!-\! \eta' \!-\! i \epsilon\Bigr)^2 \; ;
\end{eqnarray}} 
\item{There are only all $+$ vertices, which are the same as for the Feynman rules, and
$-$ vertices, which are complex conjugated;}
\item{Every 1PI $N$-point function of the Feynman rules corresponds to $2^N$ 1PI 
$N$-point functions in the Schwinger-Keldysh formalism; and}
\item{The effective field equation (\ref{EFE}) uses,
\begin{equation}
\widetilde{M}^2(x;x') = \widetilde{M}^2_{++}(x;x') + \widetilde{M}^2_{+-}(x;x') \; .
\end{equation}}
\end{itemize}
It is therefore trivial to convert our in-out result (\ref{Mren}) into the analogous
Schwinger-Keldysh result.

The next step in our program is to solve equation (\ref{EFE}) 
for one loop corrections to the plane wave mode function and to the exchange potential:
\begin{eqnarray}
\widetilde{J}(\eta,\vec{x}) = 0 & \Longrightarrow & \widetilde{\phi}(\eta,\vec{x}) = 
\Bigl\{ e^{-i k \eta} + \kappa^2 u_1(\eta,k) + O(\kappa^4)\Bigr\} 
e^{i \vec{k} \cdot \vec{x}} \; , \qquad \label{mode} \\
\widetilde{J}(\eta,\vec{x}) = \delta^3(\vec{x}) & \Longrightarrow & -
\frac1{4 \pi \Vert \vec{x} \Vert} \Bigl\{ 1 + \kappa^2 \Phi_1(\eta,\Vert \vec{x}\Vert) 
+ O(\kappa^4) \Bigr\} \; . \qquad \label{pot}
\end{eqnarray}
Although we do not need the $\delta$ counterterm (\ref{ctdelta}) to remove ultraviolet
divergences, we expect that its finite part in expression (\ref{delta}) can be chosen 
to free $u_1(\eta,k)$ of any secular enhancement. However, we also anticipate that the 
one loop correction to the potential will take the same form that was found for scalar 
corrections to gravitational potentials \cite{Park:2015kua},
\begin{equation}
\Phi_1(\eta,\vec{x}) = \frac{k_1}{a^2 \Vert \vec{x} \Vert^2} + k_2 H^2 \ln(a) 
+ k_3 H^2 \ln(a H \Vert \vec{x}\Vert) \; , \label{hunch}
\end{equation}
where $k_1$, $k_2$ and $k_3$ are constants. The term proportional to $k_1$ descends
from known effects in flat space and is anyway negligible at large distances. However,
the potentially large logarithms proportional to $k_2$ and $k_3$ are de Sitter effects 
associated with inflationary particle production. It is the gauge dependence of these
effects that we seek to establish by checking that they persist when source and 
observer effects have been included. The next steps in our program are therefore:
\begin{enumerate}
\item{Solve equation (\ref{EFE}) for the case (\ref{pot}) to verify (\ref{hunch});
and}
\item{Show that the constants $k_2$ and/or $k_3$ are nonzero when source and observer
corrections have been included.}
\end{enumerate}

It is worth mentioning other approaches to defining gauge independent correlators. The
simplest is by taking the expectation values of (necessarily nonlocal) invariant 
operators \cite{Tsamis:1989yu,Miao:2012xc,Miao:2017vly,Frob:2017apy,Frob:2017lnt,
Frob:2017gyj,Frob:2018tdd,Chataignier:2019kof}. A closely related program is the
gravitational dressing advocated by Giddings and collaborators \cite{Donnelly:2016rvo,
Giddings:2018umg,Giddings:2019hjc,Giddings:2019wmj}. Proposals have also been made for
defining invariant observables in loop quantum gravity \cite{Giesel:2017roz,
Giesel:2018opa} and in algebraic quantum field theory \cite{Brunetti:2016hgw}.
Finally, we should mention the technique of cosmological averaging 
\cite{Fanizza:2019pfp}.

Before closing we should also comment on the previous computation of $-i 
\widetilde{M}^2(x;x')$ by Boran, Kahya and Park \cite{Boran:2014xpa,Boran:2017fsx}.
Although this was a significant piece of work, it suffers from three problems. First, 
the earlier result was expressed in terms of unwieldy, de Sitter covariant differential 
operators, rather than the simple, de Sitter breaking operators $\partial^2$ and 
$\nabla^2$ that we employed in expression (\ref{Mren}). The de Sitter operators are so 
complicated that comparison is not easy but the two results do not agree. Second, Fr\"ob 
has identified a problem in the flat space correspondence limit of the earlier computation 
\cite{Frob:2017apy}, which our result avoids. Finally, using the earlier result to solve 
equation (\ref{EFE}) for case (\ref{mode}) results in one loop corrections $u_1(\eta,k)$ 
that grow like $\ln(a)$ \cite{Boran:2017cfj}, whereas it is obvious from our result 
(\ref{Mren}) that the arbitrary constant $\delta_{\rm fin}$ can be chosen to absorb any 
such enhancement.

\centerline{\bf Acknowledgements}

This work was partially supported by the Fonds de la Recherche
Scientifique -- FNRS under Grant IISN 4.4517.08 -- Theory of fundamental
interactions; by Taiwan MOST grant 108-2112-M-006-004; by the D-ITP 
consortium, a program of the Netherlands Organization for Scientific 
Research (NWO) that is funded by the Dutch Ministry of Education, 
Culture and Science (OCW); by NSF grants PHY-1806218 and PHY-1912484; 
and by the Institute for Fundamental Theory at the University of Florida.
The authors also wish to express their gratitude to NCKU for providing 
office space during this project.

\end{document}